\documentclass[12pt]{article}
\usepackage{amsmath,amsthm,amssymb}
\usepackage{algorithm,algorithmic}
\usepackage[title]{appendix}
\usepackage{color}
\usepackage{booktabs}
\usepackage{chngcntr}
\usepackage{comment}
\usepackage{dsfont}
\usepackage{endnotes}
\usepackage{enumerate}
\usepackage{float}
\usepackage[bottom]{footmisc}
\usepackage{geometry}
\usepackage{graphicx}
\usepackage[hidelinks]{hyperref}
\usepackage{indentfirst}
\usepackage[mathlines]{lineno}
\usepackage{longtable}
\usepackage{lscape}
\usepackage{mathabx}
\usepackage{makecell}
\usepackage{multibib}
\usepackage{multirow}
\usepackage{natbib}
\usepackage{pifont}
\usepackage{setspace}
\usepackage{subcaption}
\usepackage{threeparttable}
\usepackage{titling,titlesec}
\usepackage{verbatim}
\usepackage{enumitem}
\bibpunct{(}{)}{;}{a}{,}{,}
\usepackage[title]{appendix}

\newcommand{\id}{\mathrm{id}}

\usepackage{array}
\newcolumntype{L}[1]{>{\raggedright\let\newline\\\arraybackslash\hspace{0pt}}m{#1}}
\newcolumntype{C}[1]{>{\centering\let\newline\\\arraybackslash\hspace{0pt}}m{#1}}
\newcolumntype{R}[1]{>{\raggedleft\let\newline\\\arraybackslash\hspace{0pt}}m{#1}}

\floatstyle{ruled}
\newfloat{algorithm}{tbhp}{loa}
\floatname{algorithm}{Algorithm}

\titleformat{\section}{\normalfont\Large\bfseries}{\thesection}{0.5em}{}
\titlespacing*{\section} {0pt}{5pt}{3pt}
\titlespacing*{\subsection} {0pt}{5pt}{2pt}
\numberwithin{equation}{section}
\theoremstyle{plain}
\newtheorem{theorem}{Theorem}

\allowdisplaybreaks[4]
\begin{document}
\title{Generalized Causal Tree for Uplift Modeling\thanks{This work was initiated when P.\ Nandy, Y.\ Tu, K.\ Basu, and S.\ Chatterjee were at LinkedIn, and X. Yu was interning at LinkedIn. P.\ Nandy is now at Google.} \vspace{-4ex}}
\author{
\normalsize
Preetam Nandy$^1$\protect\thanks{The first two authors contributed equally.}, Xiufan Yu$^2$\protect\footnotemark[2], Wanjun Liu$^3$, Ye Tu$^4$, Kinjal Basu$^4$, Shaunak Chatterjee$^4$ \\ \normalsize $^1$LinkedIn Corporation,  $^2$University of Notre Dame, $^3$LinkedIn Corporation,  $^4$Aliveo AI Corp
}

\date{}
\maketitle{}

\pagestyle{plain}

\vspace{-3ex}
\begin{abstract}
Uplift modeling is crucial in various applications ranging from marketing and policy-making to personalized recommendations. The main objective is to learn optimal treatment allocations for a heterogeneous population. A primary line of existing work modifies the loss function of the decision tree algorithm to identify cohorts with heterogeneous treatment effects. Another line of work estimates the individual treatment effects separately for the treatment group and the control group using off-the-shelf supervised learning algorithms. The former approach that directly models the heterogeneous treatment effect is known to outperform the latter in practice. However, the existing tree-based methods are mostly limited to a single treatment and a single control use case, except for a handful of extensions to multiple discrete treatments. 
In this paper, we propose a generalization of tree-based approaches to tackle multiple discrete and continuous-valued treatments. We focus on a generalization of the well-known causal tree algorithm due to its desirable statistical properties, but our generalization technique can be applied to other tree-based approaches as well. The efficacy of our proposed method is demonstrated using experiments and real data examples. 
\end{abstract}


\section{Introduction} \label{section: intro}
Uplift modeling \citep{gutierrez2017causal,zhang2021unified} 
aims to identify individuals who benefit the most as a result of receiving a certain intervention. It has a wide range of applications, including political or marketing campaigns, 
policy-making,
ad targeting, 
personalized medicine, 
and much more. Under Rubin's framework of causal inference \citep{rubin1974estimating}, uplift modeling amounts to estimating the Conditional Average Treatment Effect (a.k.a.\ heterogeneous causal effect or individual treatment effect) for each individual based on their characteristics (or heterogeneity factors). Uplift modeling usually considers randomized control experiment data with a treatment group and a control group. The treatment group is a randomly selected population segment where each individual receives the intervention (or treatment) under consideration (e.g., medical treatment or a marketing campaign). The control group is another disjoint random segment to which the intervention is not applied. The aim is to predict the difference in the counterfactual responses when an individual receives treatment versus when the same individual receives control. These estimates are then used to decide whether the treatment should be applied. An important modification to the problem is when several treatments (possibly infinitely many) are available to us. In this case, we need to decide whether a treatment should be applied or not and which treatment (e.g., medicine dose) is the most beneficial in a given case. 

Though there is a huge literature on uplift modeling, most of the focus has been on binary treatments, that is, one treatment and one control; see \cite{zhang2021unified} for a recent literature review. 
It is of utmost importance to develop methods for uplift modeling with multiple discrete and continuous treatments, e.g., 
determining continuous weight parameters for search or recommender systems \citep{agarwal2018online}, choosing different doses of a drug \citep{jin2008principal}, choosing various thresholds for sending notifications or emails \citep{gupta2016email,liu2023quantifying}.
Recently, there are some methods that have tried to tackle the finitely many multiple-treatment \citep{rzepakowski2012decision, zhao2017uplift,guo2020survey, zhou2023offline} or continuous-treatment  \citep{oprescu2019orthogonal,bica2020estimating,wan2022gcf} scenarios.
In this paper, we propose a novel generalization of the well-known Causal Tree \citep{AtheyImbens16} algorithm, called Generalized Causal Tree (GCT), for uplift modeling with multiple discrete or continuous treatments. We chose to generalize the causal tree algorithm due to its desirable statistical properties, but our generalization technique can also be applied to other decision-tree-based approaches.

The paper is organized as follows. Section \ref{section: causal tree} introduces the problem setup and the causal tree approach. Section \ref{section: GCT} discusses the details of the GCT algorithm and an implementation-friendly variant of GCT. 
The output of GCT is user-friendly in the sense that it is (i) interpretable for understanding the treatment effect diversity and (ii) easily utilizable for optimal treatment allocations. We demonstrate the efficacy of our algorithm through empirical evaluations in Section \ref{section: simulation} and uplift analysis on two real-world datasets in Section \ref{section: realdata} before concluding with a discussion in Section \ref{section: discussion}. We end this section by highlighting some of the related works.


{\bf Related Works:} The uplift modeling problem can be decomposed into several sub-regression problems that can be solved with any supervised learning method. These types of approaches are called meta-algorithms or meta-learners \citep{KunzelEtAl19}. A sub-class of meta-learners are the so-called two-model approaches (a.k.a.\ T-learner) that build two separate predictive models for the treatment and control groups. The two-model approach naturally extends to continuous treatments (or multiple discrete treatments) by modeling the treatment group response as a function of the covariates and the treatment value. 
The treatment indicator plays a different role than the other features in the two-model approach, but there exist meta-learners where the treatment indicator is considered as an additional covariate (or feature), and the treatment effect predictions are computed from a single supervised learning model \citep{Hill11, GreenKern12}. These single-model based meta-learners are also known as S-learners, and they can also be extended to continuous treatments. 
  There are other meta-learner approaches that use special formulations for binary treatments, including X-learner \citep{KunzelEtAl19} and PW-learner \citep{CarthSchaar21}. 
  The main advantage of this approach is that state-of-the-art machine learning models (such as Random Forest and XGBoost) can be used directly for the ``uplift'' 
  prediction. However, good prediction performances for both the treatment and control group models do not guarantee good uplift predictions. One of the main reasons is that the most relevant variables for the treatment model or the control model might not be the most relevant variables for the uplift predictions. We refer to Section 5 of \cite{RadcliffeSurry11} for a simulation-based illustration along with a nicely presented list of intuitive arguments on the failure of the two-model approach. For further empirical evidence on the disadvantage of the two-model approach, see \cite{KnausEtAl20, ZaniewiczJaroszewicz13}.

Another line of work models the treatment effect directly by modifying well-known classification machine learning algorithms. For example, \citet{Guelman2014OptimalPT} focused on $k$-nearest neighbors while \citet{ZaniewiczJaroszewicz13} proposed a modification of the SVM model. 
Other methods in the literature are based on modifications of the splitting criterion in decision trees \citep{AtheyImbens16, RadcliffeSurry11, rzepakowski2012decision}.  \citet{HansotiaRukstales02} proposed a splitting criterion based on the difference in the uplift between two leaves. \citet{RadcliffeSurry11} proposed to fit a linear model with an interaction term representing the relationship between treatment and the split and defined a splitting criterion based on the significance of the interaction term. \citet{rzepakowski2012decision} described three splitting criteria based on the divergence between the treatment and control groups in the leaves. \citet{AtheyImbens16} proposed the causal tree algorithm through a treatment effect based splitting criterion. 
Most of these approaches are focused on binary uplift models, i.e., single treatment and single control use case. A natural extension to multiple treatment cases is to build binary uplift models for each treatment separately. However, this simple extension is infeasible for continuous treatments. Moreover, building separate models for each treatment is computationally and statistically inefficient. To address these issues, some authors \citep{rzepakowski2012decision, zhao2017uplift} proposed appropriate adjustments to the splitting criteria that can tackle multiple treatments simultaneously in a modified decision tree approach. Unfortunately, these extensions cannot handle continuous treatments without an ad hoc discretization of treatment values. Our main contribution is to show that there is no need for an ad hoc discretization of continuous treatment values when we can learn an optimal segmentation of the treatment values from data.

\section{Uplift Modeling using Causal Tree}
\label{section: causal tree}

We define uplift modeling as a problem of optimal treatment selection to maximize future average response in a population using randomized experiment data. 
We discuss how the causal tree algorithm 
can be used to maximize the average response in the single treatment case. Next, we point out the main disadvantages of the existing causal-tree related approaches for the multiple discrete and continuous treatment use cases.



\textbf{Problem Definition:} 
We consider a randomized experiment with control and treatment groups. The data consists of the treatment assignment $T \in \{0\}\cup \mathcal{Z}$, the response variable $Y \in \mathbb{R}$, and potential heterogeneity factors $\mathbf{X} \in \mathcal{X}$, where $\mathcal{Z}$ and $\mathcal{X}$ denote the set of treatment values in the treatment group and the set of all values taken by $\mathbf{X}$ respectively. Without loss of generality, we assume that each member in the control group received $T = 0$, and each member in the treatment group is assigned a treatment value $Z \in \mathcal{Z} \subseteq \mathbb{R}\setminus\{0\}$ generated from a (discrete or continuous) probability distribution $p_{Z}(\cdot)$ on $\mathcal{Z}$. 
For the single treatment case, $p_{Z}(\cdot)$ is a point mass distribution on $\mathcal{Z} = \{1\}$ (without loss of generality). In this paper, we do not discuss the choice of $\mathbf{X}$ but focus on exploiting the heterogeneity of the conditional treatment effect to maximize the average response for a given $\mathbf{X}$. The set of potential heterogeneity factors $\mathbf{X}$ are assumed to be pre-treatment covariates (i.e., $\mathbf{X}$ is independent of $T$).  


Let $\mu(t, \mathbf{x}) = E[Y(T = t) \mid X = \mathbf{x}]$ denote the average counterfactual response for individuals with characteristics (or heterogeneity factors) $\mathbf{X} = \mathbf{x}$ under the treatment 
$T = t$. We further define $\tau(z, \mathbf{x}) =  \mu(z, \mathbf{x}) - \mu(0, \mathbf{x})$ to be the conditional treatment effect for $\mathbf{X} = \mathbf{x}$ and $z \in \mathcal{Z}$. For single treatment case, we denote the individual treatment effect by $\tau(\mathbf{x})$.


\textbf{Causal Tree:} The single treatment causal tree algorithm of \citet{AtheyImbens16} fits a decision tree model for estimating conditional treatment effect $\tau(\mathbf{x})$ with a splitting criterion that maximizes the sum of squared treatment effects in the leaves (i.e., nodes without children) while penalizing for higher estimation variances. The output of the causal tree algorithm is a partition (i.e., a disjoint and exhaustive collection) $\Pi_1, \ldots, \Pi_K$ of $\mathcal{X}$ and the estimated causal effects $\hat{\tau}(\Pi_1), \ldots, \hat{\tau}(\Pi_K)$ in those cohorts. Based on the estimated causal effects, one can choose treatment or control for each cohort for future assignments, where the within cohort treatment effects are assumed to be homogeneous (or statistically indistinguishable). \citet{TuEtAl21} used the single treatment causal tree algorithm in a multi-treatment uplift modeling setting by building separate trees for each treatment and then merging the trees to obtain a finer partition  $\Pi_1', \ldots, \Pi_K'$ of $\mathcal{X}$. The authors also described a procedure to obtain cohort-specific treatment effect estimates $\tau(z, \Pi_i')$ for all $z \in \mathcal{Z}$ using the estimated treatment effects in the separately built trees. These estimates can be used to maximize the future total effect by assigning 
\begin{align*}
v(\mathbf{x}) = \left\{ \begin{array}{ll}
   \underset{ z \in \mathcal{Z}}{\arg \max}~ \hat{\tau}(z, \Pi_i')  & \text{if } \underset{z \in \mathcal{Z}}{\max}~ \hat{\tau}(z, \Pi_i') > 0, \\
     0 & \text{otherwise}
\end{array} \right.
\end{align*}
to each $\Pi_i'$ for $i = 1,\ldots, K$.

{\bf Sample Inefficiency}: This approach can suffer from sample inefficiency by ignoring the presence of similarities between treatments. For example, suppose treatments $T=1$ and $T=2$ are identical, then there is no need to learn separate causal trees for $T=1$ and $T=2$ by splitting the data corresponding to $T \in \{1, 2\}$. Of course, we may not have prior knowledge of treatment similarities. Still, we can be more sample efficient by learning treatment cohorts (i.e., a similarity-based partition of the treatment set $\mathcal{Z}$) jointly with the $\mathbf{X}$ cohorts. Moreover, the partitioning of the treatment set is a necessity when $\mathcal{Z}$ contains infinitely many values (e.g., continuous treatment), as building separate trees is not possible anymore.

{\bf Continuous Treatment}: 
Existing continuous treatment version of the causal tree or causal forest \citep{stefan2015forest} algorithm defines the treatment effect to be $\tau(\mathbf{x}) := \frac{\partial}{\partial t} \mu(t, \mathbf{x})$, assuming the linearity of $\mu(t, \mathbf{x})$ with respect to $t$. The linearity assumption is crucial for having constant $\tau(\mathbf{x})$ for all values of $t$, and in this case $\tau(\mathbf{x})$ can be interpreted as the change in the average response for changing the treatment to $t$ to $t+1$. The linearity is a strong assumption here, and consequently, the optimal treatment will always be either the maximum or the minimum treatment value. Therefore, we avoid the use of such a technique for uplift modeling. 
In the following section, we propose the GCT algorithm that produces a partition $\Pi_1, \ldots, \Pi_K$ of $\mathcal{X}$, a partition $\Gamma_1, \ldots, \Gamma_K$ of $\mathcal{Z}$, and the estimates of causal effects $\hat{\tau}(\Gamma_j, \Pi_i)$ for each $(\Gamma_j, \Pi_i)$ pair. We emphasize that GCT groups similar treatments together based on homogeneity of the treatment effects jointly with the identification of homogeneous $\mathbf{X}$-cohorts from data. This is different from applying a multiple treatment uplift modeling after discretizing continuous treatment values based on quantiles or ad hoc cut points.

\textbf{Assumptions and Limitations:} Under Rubin's potential outcome framework \citep{rubin1974estimating}, we require the following standard assumptions of the individual treatment effects (ITE) literature \citep{BicaEtAl20, ImaiDYK04, Imbens00, SchulamSaria17,  ZhangEtAl20}: (i) unconfoundedness and (ii) the Stable Unit Treatment Value assumption (SUTVA). The unconfoundedness assumption states that the treatment variable and the potential outcomes are conditionally independent given the pre-treatment covariates. SUTVA ensures that the potential outcome of an experiment unit is unaffected by the treatment assignment of the other units. Additionally, we assume completely randomized treatment (i.e., the independence of the treatment variable and the pre-treatment covariates) and the existence of a control dataset (i.e., $P(T = 0) >0$). These assumptions are more common in the uplift modeling literature than in the ITE estimation literature. It is straightforward to remove the completely randomized treatment assumption by applying, for example, inverse propensity score weighting techniques. However, the proposed methodologies are only applicable in settings with a control dataset and a treatment dataset with different treatment values. 



\section{Generalized Causal Tree (GCT)}\label{section: GCT}

We consider data from a randomized experiment with a control group corresponding to $\{T = 0\}$ and a treatment group corresponding to $\{T \in \mathcal{Z}\}$, as described in Section \ref{section: causal tree}. We define $W = \mathds{1}_{\{T \in \mathcal{Z}\}}$ to be the treatment group indicator and $Z_1 = (T \mid W = 1)$ to be the treatment value in the treatment group. We use the subscript ``1'' in $Z_1$ to emphasize the fact that $Z_1$ is defined only for the treatment group $\{W=1\}$. Let 
$\mathbb{S}_0^{tr} = \{(\mathbf{x}_i^{(0)}, y_i^{(0)})\}_{i=1}^{N_0}$ and $\mathbb{S}_1^{tr} = \{(\mathbf{x}_i^{(1)}, y_i^{(1)}, z_i^{(1)})\}_{i=1}^{N_1}$ denote the data corresponding to the control group $\{W=0\}$ and the treatment group $\{W=1\}$ respectively containing pre-treatment covariates $\mathbf{x}_i^{(w)}$ and response $y_i^{(w)}$ for $w = 0,1$, and treatment value $z_i^{(1)}$. Note that we can fit a single treatment causal tree to $\mathbb{S}^{tr}$ based on the binary treatment indicator $W$ (by ignoring the $Z_1$ values). We generalize this by adding the use of $Z_1$ in the causal tree splitting criterion defined by  \eqref{eq: objective function}.

Although we focus on the causal tree algorithm, our strategy works for any decision tree-based uplift modeling method. To exhibit this, we use a generic form of the objective function of the causal tree algorithm. At every step, the single treatment causal tree algorithm based on the treatment indicator $W$ splits a leaf node in the current decision tree $\mathcal{D}$ to increase (if possible) the value of an objective function of the form
\begin{align}\label{eq: objective function}
\mathcal{O}(\mathcal{D}, \mathbb{S}^{tr}) = \sum\nolimits_{i=1}^{k} g\left(h(\mathbb{S}^{tr}_1(\Pi_i)),~ h(\mathbb{S}^{tr}_0(\Pi_i)\right),
\end{align}
where $g(\cdot,\cdot)$ measures the utility of a leaf node (i.e., the square of estimated treatment effect minus the square of its estimated standard error), $h(\cdot)$ is a set of summary statistics (i.e., count, mean and variance) on responses conditioned on $\mathbb{S}^{tr}_w(\Pi_i) \subseteq \mathbb{S}^{tr}_w$, $\{\Pi_1,\ldots, \Pi_k\}$ is a partition of $\mathcal{X}$ defined by the leaf nodes in $\mathcal{D}$, and $\mathbb{S}^{tr}_w(\Pi_i) = \{(\mathbf{x}_i^{(w)}, y_i^{(w)}): 1\leq i \leq N_w,~ \mathbf{x}_i^{(w)} \in \Pi_i\}$ for $w = 0,1$.
Now, we modify the objective function \eqref{eq: objective function} to allow splitting the treatment data based on both $\mathbf{X}$ and $Z$.
\begin{align}\label{eq: modified objective function}
\mathcal{O}^*(\mathcal{D}, \mathbb{S}^{tr}) = \sum\nolimits_{i=1}^{k} g\left(h(\mathbb{S}^{tr}_1(\Lambda_i)),~ h(\mathbb{S}^{tr}_0(\Pi_i)\right),
\end{align}
where $\mathbb{S}^{tr}_1(\Lambda_i) = \{(\mathbf{x}_i^{(1)}, y_i^{(1)}, z_i^{(1)}) : 1\leq i \leq N_1,~ (\mathbf{x}_i^{(1)},z_i^{(1)})\in \Lambda_i\}.$
We call this causal tree algorithm with the modified objective function \eqref{eq: modified objective function} the basic \emph{Generalized Causal Tree} (GCT) algorithm. The output of basic GCT is a partition $\mathcal{P}_{\mathcal{X}, \mathcal{Z}} = \{\Lambda_i\}_{i=1}^{K}$ of $\mathcal{X} \times \mathcal{Z}$, and the corresponding estimates $\hat{\tau}(\Lambda_i)$. Following the ``honest'' approach \cite{AtheyImbens16}, we assume the existence of an additional dataset $\mathbb{S}^{est} := \{\mathbb{S}_0^{est}, \mathbb{S}_1^{est}\}$ for estimating treatment effects in leaf nodes. Hence, we have 
$$\hat{\tau}(\Lambda_i) = 
\hat{E}[Y \mid \mathbb{S}_1^{est}(\Lambda_i)] - \hat{E}[Y \mid \mathbb{S}_0^{est}(\Pi_i)],$$
i.e., the difference between the empirical average responses in $\mathbb{S}_1^{est}(\Lambda_i)$ and $\mathbb{S}_0^{est}(\Pi_i)$. In the special case where $Z_1$ is not chosen by GCT for partitioning (because $Z_1$ is not a detectable heterogeneity factor), we have $\Lambda_i = \Pi_i \times \mathcal{Z}$ and $\hat{\tau}(\Lambda_i) = \hat{\tau}(\Pi_i) = 
\hat{E}[Y \mid \mathbb{S}_1^{est}(\Pi_i)] - \hat{E}[Y \mid \mathbb{S}_0^{est}(\Pi_i)]$.

There are two significant obstacles to using the basic GCT algorithm in practice. The first issue is the moderately high implementation cost, as one needs to modify the splitting criterion in an existing implementation of the single treatment causal tree algorithm. The second issue is that the $\mathbf{X}$-cohorts $\{\Pi_i \subseteq \mathcal{X} : (\Pi_i, \Gamma_i) \in \mathcal{P}_{\mathcal{X}, \mathcal{Z}},~\Gamma_i \subseteq \mathcal{Z}\}$ are not disjoint sets and hence they do not form a partition of $\mathcal{X}$. This makes it computationally expensive to identify an optimal treatment for each member in a population, especially in large-scale web applications with billions of members \citep{TuEtAl21}. To overcome these obstacles, we propose an implementation-friendly modification of basic GCT.
Then we address the second issue by making a user-friendly transformation of the output of basic GCT.

\subsection{Implementation-friendly Objective}\label{subsection: implementation-friendly modification}

We propose a modification of the objective function \eqref{eq: modified objective function} such that an existing implementation of the single treatment causal tree algorithm can be reused directly. We wish to learn a causal tree based on $W$ while using $\mathbf{X}$ and $Z_1$ jointly as heterogeneity factors. This is not possible without extending the definition of $Z_1$ to the control group. To this end, we assign a $Z$-value to each data point in the control group by randomly selecting a value from the distribution $p_{Z_1}(\cdot)$ of $Z_1$. We define $Z$ as
\begin{align}\label{eq: modified Z}
Z = Z_1 \times \mathds{1}_{\{W = 1\}} + Z_0 \times \mathds{1}_{\{W = 0\}}
\end{align}
where $Z_0$ is independently and identically distributed as $Z_1$. In practice, we sample from the empirical distribution of $Z_1$ if the actual distribution is unknown. By choosing the distribution of $Z_0$ the same as the distribution of $Z_1$, we make sure the same proportion of samples (on average) in the treatment and control group in each leaf node. Based on the definition of $Z$ as in \eqref{eq: modified Z}, we rewrite the training data as $\mathbb{S}^{tr} = \{(\mathbf{x}_i, y_i, z_i, w_i)\}_{i=1}^{N}$ where $N = N_0 + N_1$. Now we propose the implementation-friendly objective function
\begin{align}\label{eq: implementation-friendly objective function}
\mathcal{O}^*(\mathcal{D}, \mathbb{S}^{tr}) = \sum\nolimits_{i=1}^{k} g(h(\mathbb{S}^{tr}_1(\Lambda_i)), h(\mathbb{S}^{tr}_0(\Lambda_i)),
\end{align}
where $\mathbb{S}^{tr}_w(\Lambda_i) = \{(\mathbf{x}_i, y_i, z_i, w_i) : 1\leq i \leq N, w_i = w, (\mathbf{x}_i, z_i) \in \Lambda_i\}$ for $w$ = $0,1$. Note that the difference between GCT based on Eq.\eqref{eq: implementation-friendly objective function} and GCT based on Eq.\eqref{eq: modified objective function} is that the former splits both treatment data and control data based on a constraint on $Z$ defined in Eq.\ \eqref{eq: modified Z} while the latter keeps the control data unchanged for splits based on $Z_1 = (T \mid W = 1)$.

Note that we reduce the effective sample size corresponding to $\{W = 0\}$ through the artificially induced randomness of $Z_0$ in Eq. \eqref{eq: modified Z}. This could be a concern in situations where collecting data corresponding to $\{W = 0\}$ is expensive. However, this is not a concern in many cases, including web-scale problems where there is no shortage of data corresponding to $\{W = 0\}$ (the baseline model). Typically, we have a control group at least as large as the treatment group. In the case of equal-sized treatment and control groups, GCT based on Eq.\ \eqref{eq: implementation-friendly objective function} would have identical sample size distributions for the treatment group and control group in each $\mathcal{P}_{\mathcal{X}, \mathcal{Z}}$ cohort. This implies the same level of uncertainties in the average treatment response and average control response in those groups.

 
\textbf{Additional Notation:}
Consider a decision tree $\mathcal{D}$ based on a features $\mathbf{X}$. Recall that $\mathcal{D}$ defines a partition of $\mathcal{X}$ based on its leaf nodes. 
We assign a unique identifier $id(\ell, \mathcal{D})$ to each leaf $\ell$. We represent $\mathcal{D}$ as a full binary tree 
where each non-leaf node $a$ contains (i) a pointer to the left child node $left(a)$, (ii) a pointer to the right child node $right(a)$, and (iii) a cohort definition $\pi(a) \subseteq \mathcal{X}$ with $\pi(a) := \mathcal{X}$ for the root node $a$.
A non-leaf node $a$ is a unique parent of $left(a)$ and $right(a)$. If we decide to split a node $a$ based on a continuous feature $X_r \in \mathbf{X}$ and at the value $s$, then we have $\pi(left(a)) = \{(x_r, \mathbf{x}_{-r}) \in \mathcal{X}_r \times \mathcal{X}_{-r} : x_r \leq s\}$ and $\pi(right(a)) = \{(x_r, \mathbf{x}_{-r}) \in \mathcal{X}_r \times \mathcal{X}_{-r} : x_r > s\}$, where $\mathcal{X}_r$ and $\mathcal{X}_{-r}$ denote the sets of values taken by $X_r$ and $\mathbf{X}\setminus \{X_r\}$ respectively. We denote the parent of $a$ in $\mathcal{D}$ by $parent(a, \mathcal{D})$. A sequence of nodes $a_0,\ldots,a_k$ is an ancestral path $a_0$ to $a_k$ in $\mathcal{D}$ if for each $i \in \{1,\ldots,k\}$, $a_{i-1} = parent(a_i, \mathcal{D})$. Let $\{\ell_i\}_{i=1}^{K}$ be all leaf nodes of $\mathcal{D}$. For each leaf $\ell_i$, we define
$
\Pi_i := \Pi_{\mathbf{X}}(\ell_i, \mathcal{D}) := \cap_{r = 0}^{k_i}\pi(a_r, \mathcal{D})),
$
where $a_0,\ldots,a_{k_i}$ is the unique ancestral path from the root $a_0$ to $\ell_i = a_{k_i}$. Then $\mathcal{P}_{\mathcal{X}} := \{\Pi_i\}_{i=1}^{K}$ forms a partition of $\mathcal{X}$.


A post-order traversal on a binary tree returns an ordering of the node through a recursive depth-first traversal that first visits the left sub-tree, then the right sub-tree, and finally the root. For any $\mathbf{X}' \subseteq \mathbf{X}$, we denote the coordinate-wise projection of $\Pi_{\mathbf{X}}(a, \mathcal{D})$ onto $\mathcal{X}'$ (the set of all possible $\mathbf{X}'$ values) by $\Pi_{\mathbf{X}'}(a, \mathcal{D})$ (i.e., obtained by removing all coordinates corresponding to $\mathbf{X}\setminus \mathbf{X}'$). For convenience, we use the same notation to denote a singleton set $\{i\}$ and its element $i$.

\begin{figure}
    \centering
    \includegraphics[width=0.7\textwidth]{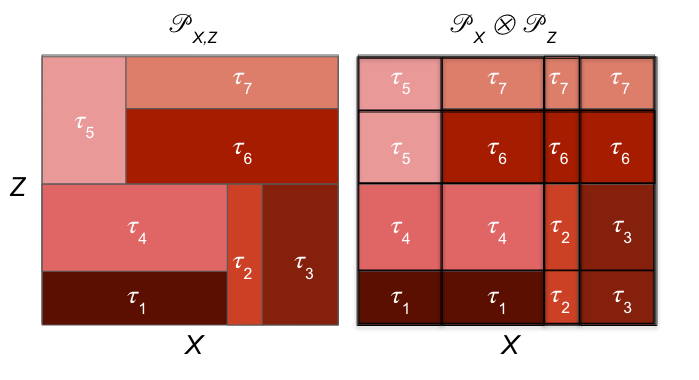}
    \caption{\small GCT output transformation visualization}
    \label{fig: partition}
\end{figure}


\subsection{User-friendly Output}\label{subsection: user-friendly output}
We transform the output of basic GCT $\mathcal{P}_{\mathcal{X}, \mathcal{Z}}$ to a cross-product $\mathcal{P}_{\mathcal{X}} \times \mathcal{P}_{\mathcal{Z}} = \{\Pi_1,\ldots, \Pi_{K_{\mathcal{X}}}\} \times \{\Gamma_1,\ldots, \Gamma_{K_{\mathcal{X}}}\}$ of partitions of $\mathcal{X}$ and partitions of $\mathcal{Z}$ and provide corresponding treatment effect estimates, i.e., $\{\hat{\tau}(\Pi_i, \Gamma_j) : \Pi_i \in \mathcal{P}_{\mathcal{X}},~ \Gamma_j \in \mathcal{P}_{\mathcal{Z}}\}$. The transformed output is much more user-friendly because the cross-product representation of cohorts makes it easy to find an optimal treatment set for each $\Pi_i \in \mathcal{P}_{\mathcal{X}}$ by setting $\Gamma_{i^*} = \underset{\Gamma_j \in \mathcal{P}_{\mathcal{Z}}}{\text{argmax}}~ \hat{\tau}(\Pi_i, \Gamma_j)$ whenever $\underset{\Gamma_j \in \mathcal{P}_{\mathcal{Z}}}{\max}~ \hat{\tau}(\Pi_i, \Gamma_j) > 0$. Then for any individual with features $\mathbf{x} \in \Pi_i$ an optimal treatment can be chosen from the conditional distribution $p_Z(\cdot \mid \Gamma_{i^*})$ of $Z$ given $\{Z \in \Gamma_{i^*}\}$.

We use tree-modification procedure (Algorithm \ref{alg: remove nodes}) to obtain $\mathcal{P}_{\mathcal{X}} \times \mathcal{P}_{\mathcal{Z}}$ and the corresponding treatment effects via Algorithm \ref{alg: generalized causal tree}. Before that, we illustrate the main idea with the toy example depicted in Figure \ref{fig: partition}. We consider a continuous-valued $Z$ and an one-dimensional continuous-valued $X$. The left sub-figure represents estimated causal effects in a basic GCT as piecewise constant functions on $\mathcal{P}_{X, Z}$, which is a collection of disjoint and exhaustive axis-aligned rectangles. The right sub-figure represents $\mathcal{P}_X \times \mathcal{P}_Z$ as a minimal (in terms of the total number of rectangles) finer partition of $\mathcal{X} \times \mathcal{Z}$ that is a cross-product of a partition of $\mathcal{X}$ (corresponding to the vertical lines in the sub-figure) and a partition of $\mathcal{Z}$ (corresponding to the horizontal lines in the sub-figure). The (color-coded) treatment effects in these finer partitions are borrowed from the treatment effects in $\mathcal{P}_{X, Z}$.


Note that the $X$-cohorts can be obtained by projecting each cohort $\Lambda \in \mathcal{P}_{X,Z}$ onto $\mathcal{X}$ and by splitting the resulting non-disjoint sets into disjoint sets appropriately. The splitting steps can be done efficiently by exploiting the underlying decision tree structure. We present this idea formally in Algorithm \ref{alg: remove nodes}, which allows us to remove a set of features $\mathbf{X}'$ from a decision tree $\mathcal{D}_{in}$ such that (1) the output is a decision tree $\mathcal{D}_{out}$ based on the feature set $\mathbf{X}\setminus \mathbf{X}'$, and (2) for each leaf node $\ell$ in $\mathcal{D}_{out}$, there exists a leaf node $\ell'$ in $\mathcal{D}_{in}$ such that $\Pi_{\mathbf{X} \setminus \mathbf{X}'}(\ell, \mathcal{D}_{out}) \subseteq \Pi_{\mathbf{X} \setminus \mathbf{X}'}(\ell', \mathcal{D}_{in}),$ where $\Pi_{\mathbf{X} \setminus \mathbf{X}'}(\cdot, \mathcal{D}_{in})$ denotes a cohort of $\mathcal{D}_{in}$ projected onto $\mathcal{X} \setminus \mathcal{X}'$.
Additionally, 
For each leaf $\ell$ in $\mathcal{D}_{out}$, we facilitate the identification of the set of super-cohort leaves in $\mathcal{D}_{in}$, defined as
\begin{align}\label{eq: super-cohot}
\{ \ell' \in \mathcal{D}_{in} : \Pi_{\mathbf{X} \setminus \mathbf{X}'}(\ell, \mathcal{D}_{out}) \subseteq \Pi_{\mathbf{X} \setminus \mathbf{X}'}(\ell', \mathcal{D}_{in}) \}.
\end{align}

\begin{algorithm}[!hbt]
    \caption{Removing features from a decision tree}
    \label{alg: remove nodes}
    \begin{algorithmic}[1]
        \REQUIRE a decision tree $\mathcal{D}$, a set features $\mathbf{X}'$
        \STATE Let $a_1, \ldots, a_n$ be the ordering of the nodes from a post-order traversal in $\mathcal{D}$;
        \FOR {$i = 1, \ldots, n$} 
            \IF {$a_i$ is a non-leaf and $\pi(left(a))$ (or $\pi(right(a)$) contains a constraint involving a feature in $\mathbf{X}'$,}
            \STATE merge $left(a_i)$ and $right(a_i)$ as follows:
                \IF { both $left(a_i)$, $right(a_i)$ are leaves}
                    \STATE \texttt{MergeLeaves}($\mathcal{D}$, $a_i$);
                \ELSIF {one of $left(a_i)$, $right(a_i)$ is a leaf}
                    \STATE \texttt{MergeLeafAndNonLeaf}($\mathcal{D}$, $a_i$);
                \ELSE
                    \STATE \texttt{MergeNonLeaves}($\mathcal{D}$, $a_i$);
                \ENDIF
            \ENDIF
         \ENDFOR
        \STATE Return $\mathcal{D}$
    \end{algorithmic}
\end{algorithm}

These sets of super-cohort leaves will be used in Algorithm \ref{alg: generalized causal tree} for computing treatment effects. 
The main idea of Algorithm \ref{alg: remove nodes} is to 
sequentially visit each node $a$ via a post-order traversal and remove its children $(left(a, \mathcal{D}), right(a, \mathcal{D}))$ whenever the split is defined by a feature in $\mathbf{X}'$, and merge the sub-trees below the children with the parent node $a$. The merging operation involves updating the children pointers and removing all nodes corresponding to a conflicting cohort definition. Moreover, we keep track of the super-cohort sets defined in \eqref{eq: super-cohot} by passing the identifier information of the deleted nodes and the nodes that has lost their leaf status due to the merging. 

\begin{figure}[htb]
    \centering
    
    \begin{subfigure}{0.48\textwidth}
        \begin{subfigure}{0.55\textwidth}
        \centering
        \includegraphics[width=0.9\linewidth]{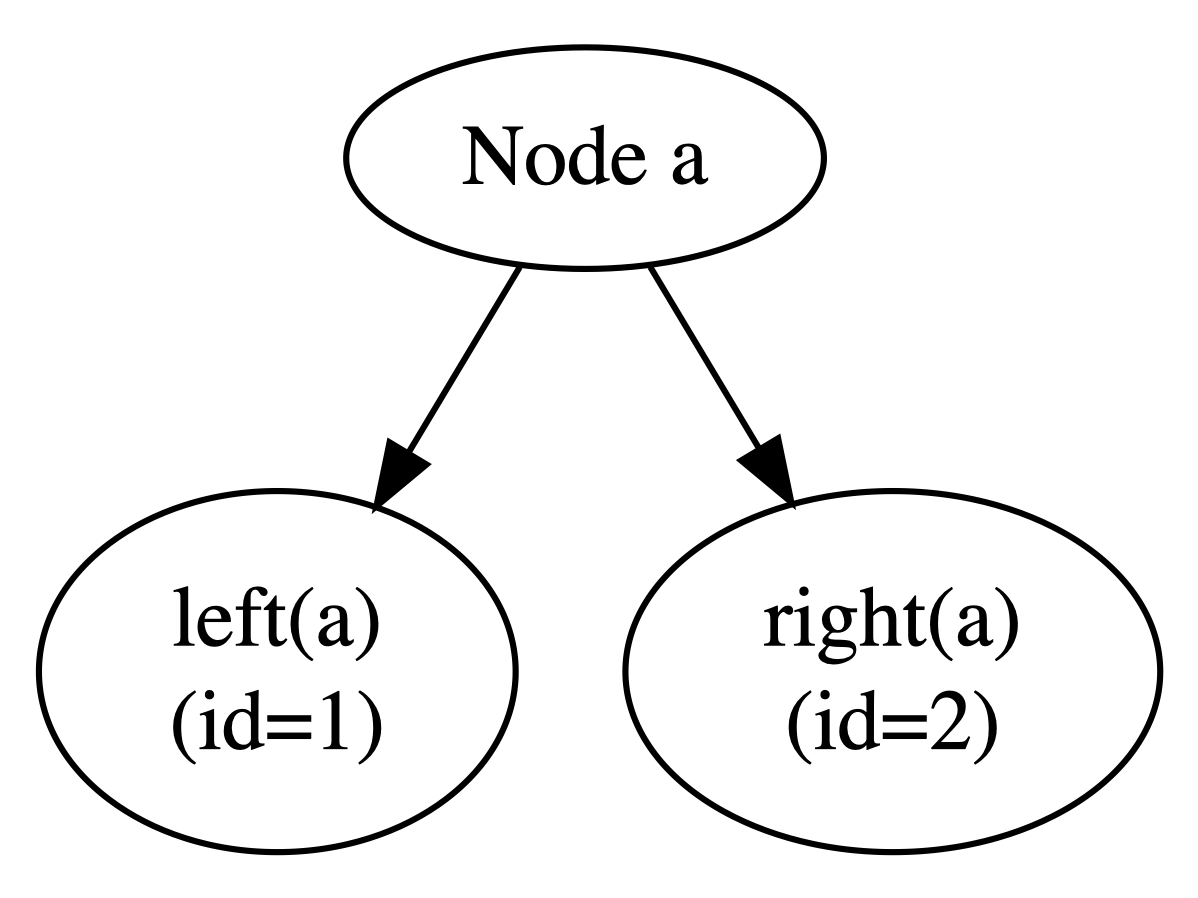}
        \renewcommand\thesubfigure{\alph{subfigure}1}
        \caption{Before merging}
        \end{subfigure}
        \begin{subfigure}{0.4\textwidth}
        \centering
        \addtocounter{subfigure}{-1}
        \includegraphics[width=0.7\linewidth]{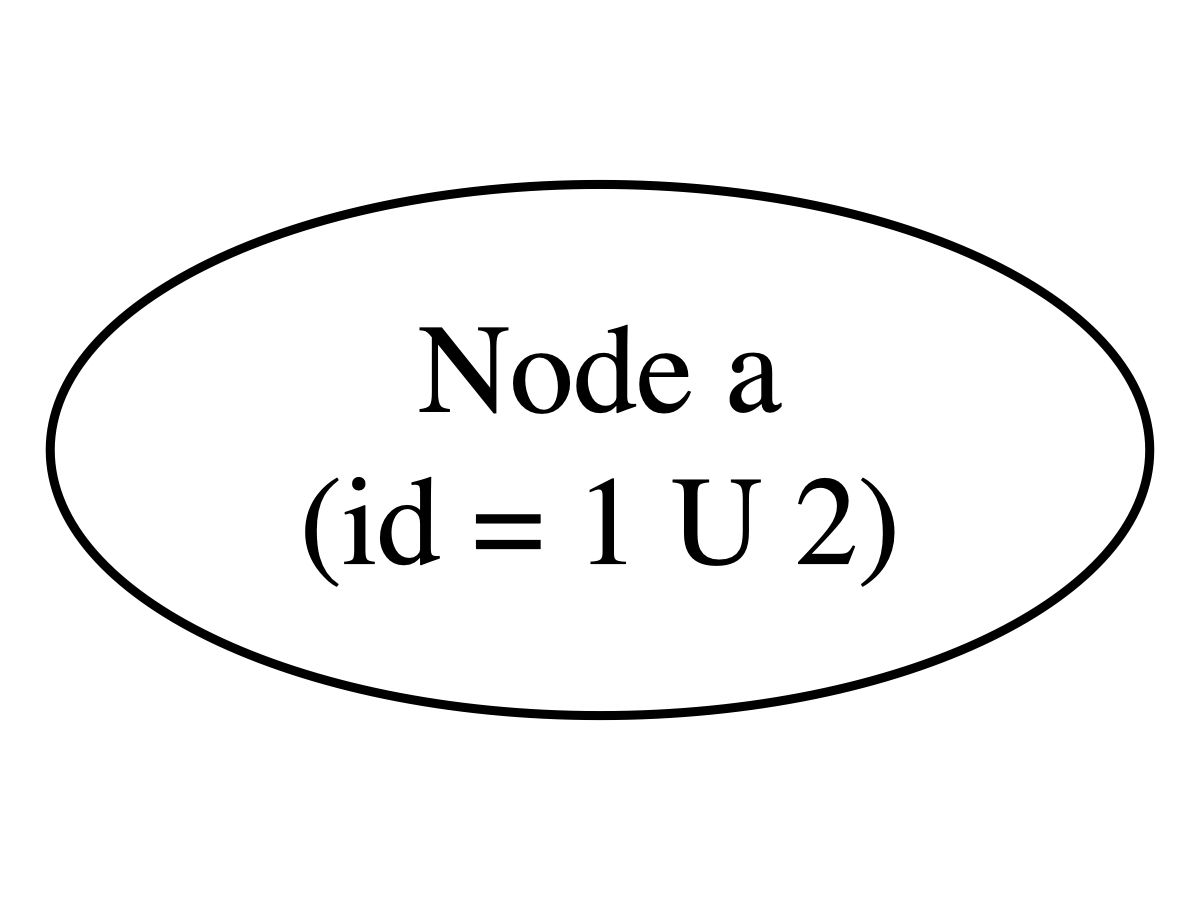}
        \renewcommand\thesubfigure{\alph{subfigure}2}
        \caption{After merging}
        \end{subfigure}
        \addtocounter{subfigure}{-1}
        \captionsetup{justification=centering}
        \caption{Example of merging leaf children (Algorithm \ref{alg: merging leaves})} \label{fig: MergeLeafChildren}
    \end{subfigure}
    \begin{subfigure}{0.48\textwidth}
        \begin{subfigure}{0.52\textwidth}
        \centering
        \renewcommand\thesubfigure{\alph{subfigure}1}
          \includegraphics[width=\linewidth]{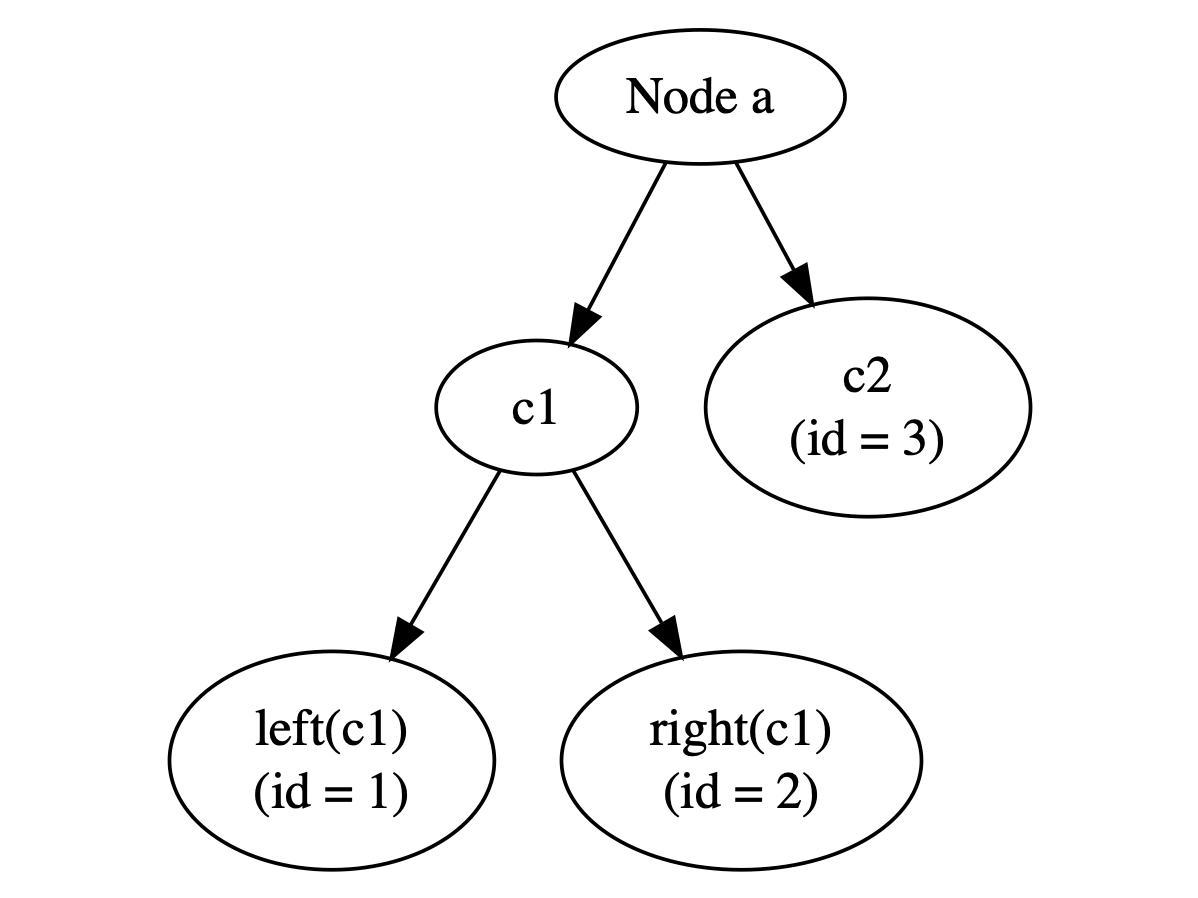}
          \caption{Before merging}
        \end{subfigure}
        \begin{subfigure}{0.45\textwidth}
        \centering
         \addtocounter{subfigure}{-1}
        \renewcommand\thesubfigure{\alph{subfigure}2}
          \includegraphics[width=0.9\linewidth]{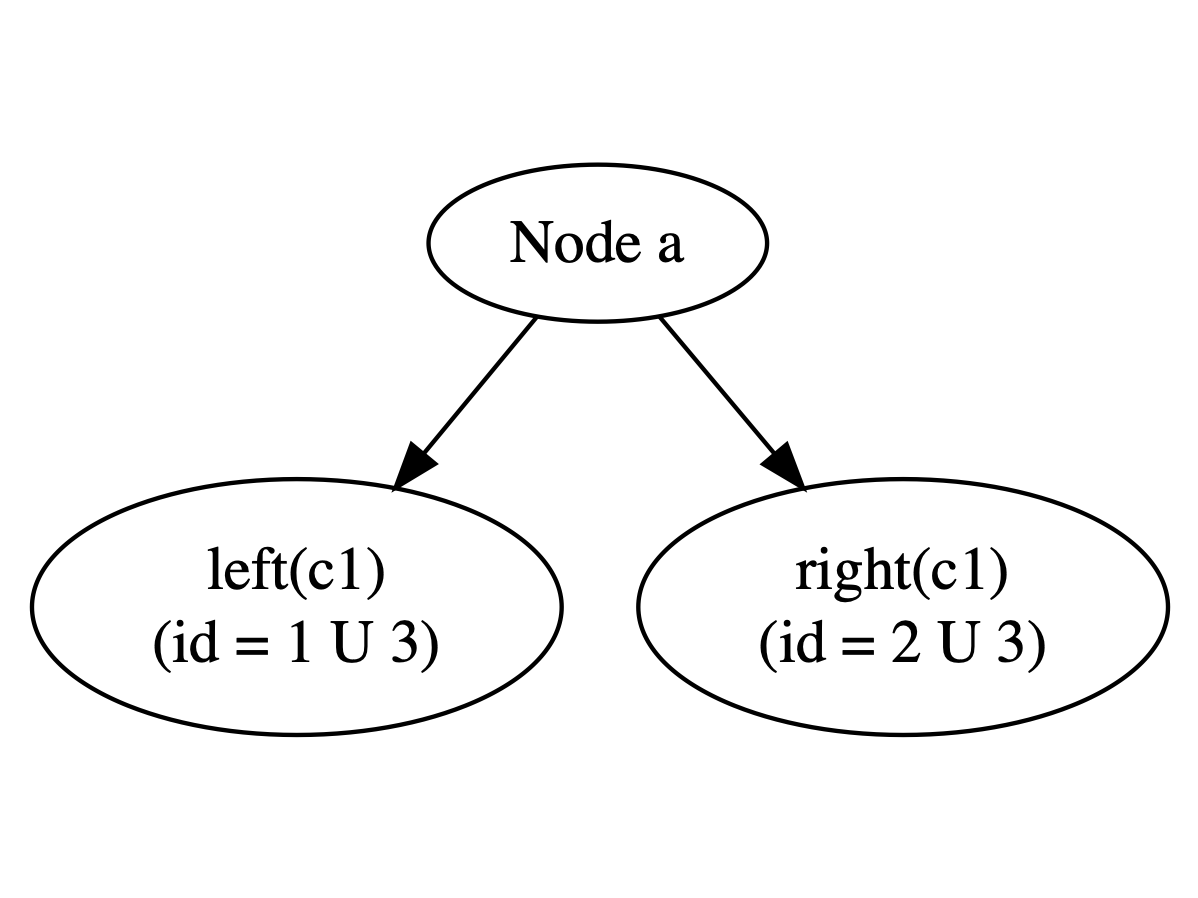}
          \caption{After merging}
        \end{subfigure}
        \addtocounter{subfigure}{-1}
        \captionsetup{justification=centering}
        \caption{Example of merging leaf and non-leaf children (Algorithm \ref{alg: merging leaf and non-leaf})} \label{fig: MergeLeafAndNonLeafChildren}
    \end{subfigure}
    
    \bigskip
    
    \begin{subfigure}{\textwidth}
            \begin{subfigure}{0.45\textwidth}
          \centering
          \renewcommand\thesubfigure{\alph{subfigure}1}
          \includegraphics[width=\linewidth]{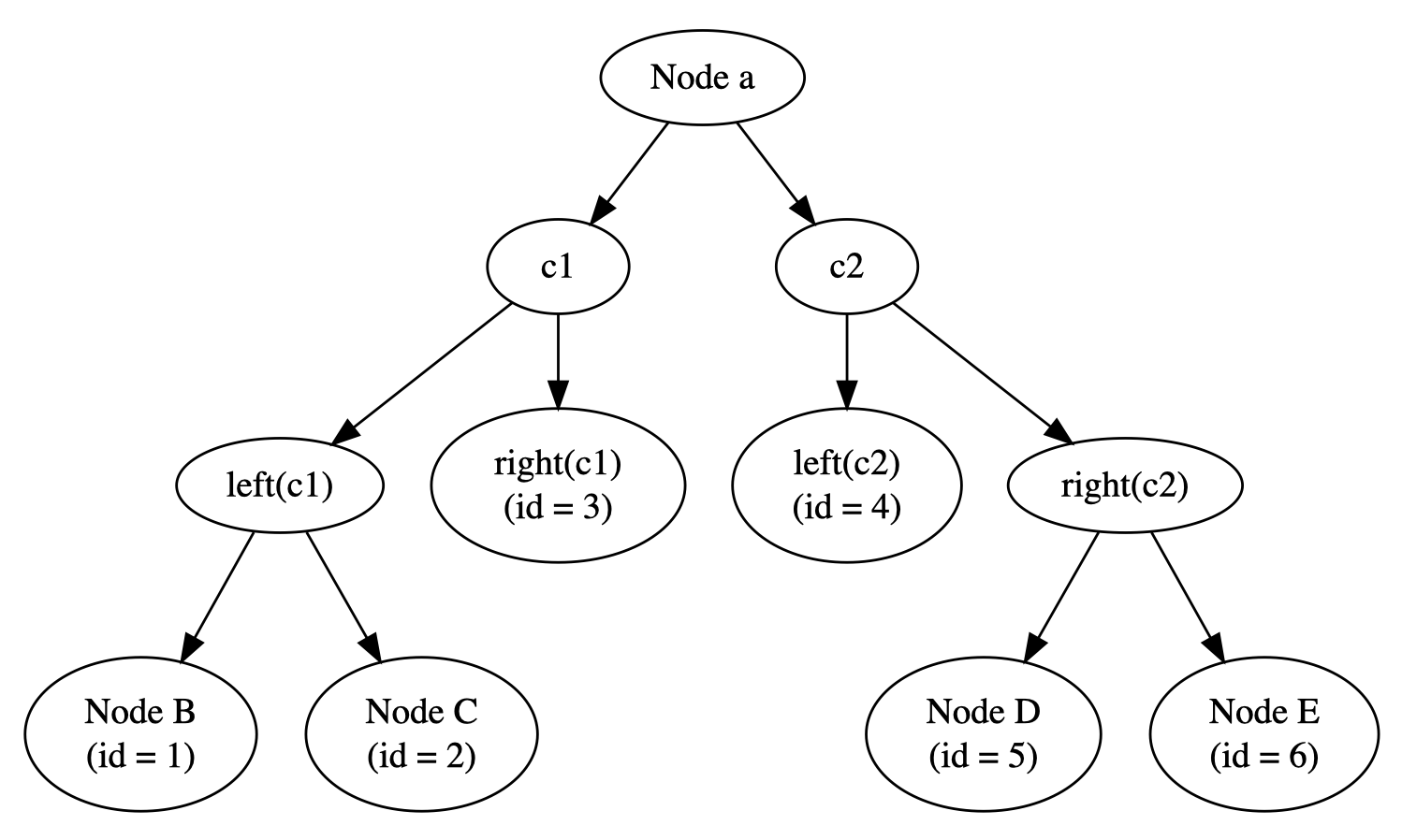}
          \caption{Before merging}
        \end{subfigure}
        \begin{subfigure}{0.5\textwidth}
          \centering
          \addtocounter{subfigure}{-1}
          \renewcommand\thesubfigure{\alph{subfigure}2}
          \includegraphics[width=\linewidth]{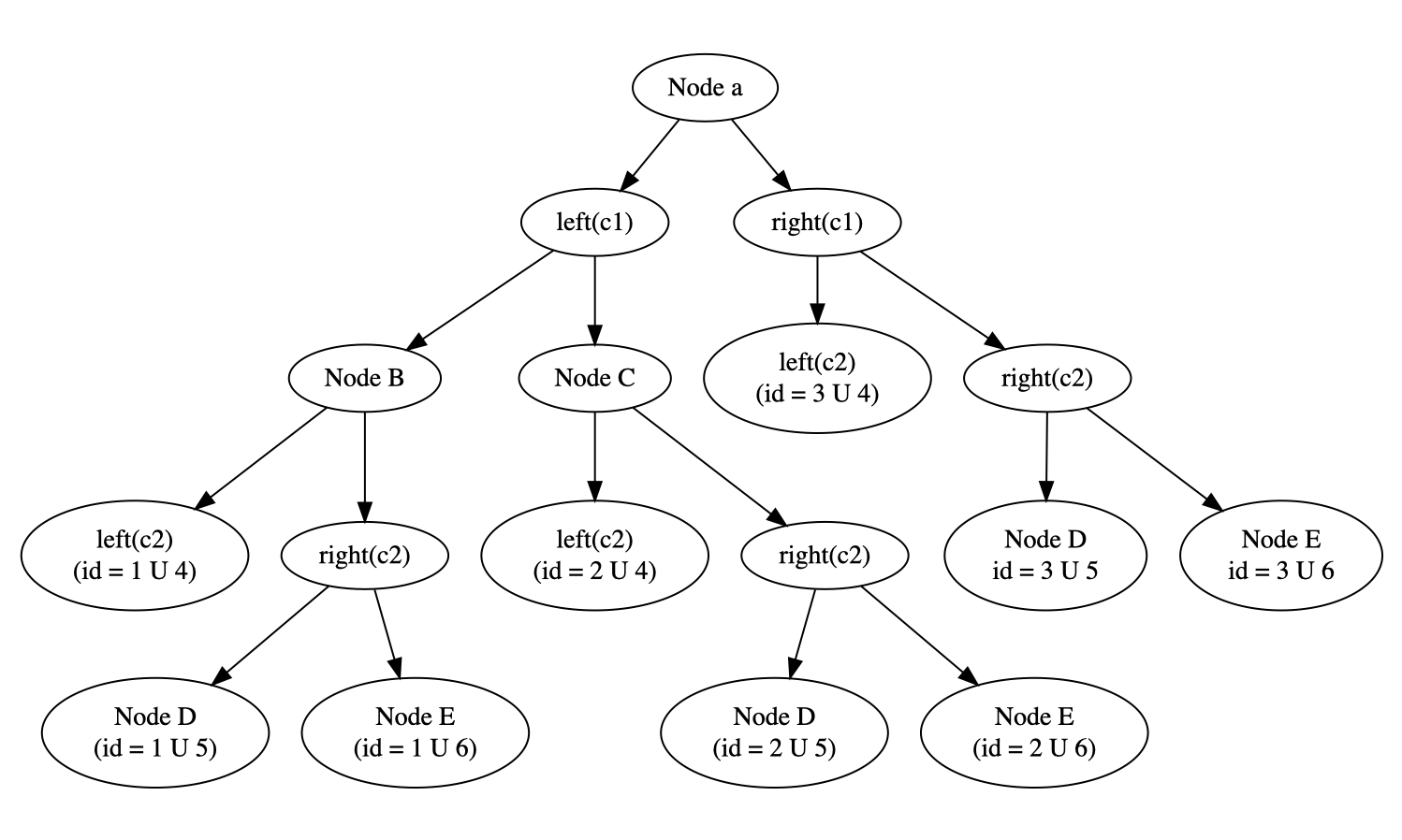}
          \caption{After merging}
        \end{subfigure}
        \addtocounter{subfigure}{-1}
        \captionsetup{justification=centering}
        \caption{Example of merging non-leaf children (Algorithm \ref{alg: non-leaf nodes})} \label{fig: MergeNonLeafChildren}
    \end{subfigure}
    \caption{Examples of Algorithms \ref{alg: merging leaves}-\ref{alg: non-leaf nodes}} \label{fig: eg-Algorithms234}
\end{figure}

The removing algorithm sequentially visits each node via a post-order traversal and removes a split from the tree whenever the split is associated with the targeted feature. Once a split is identified as a to-be-removed branch, the removal process not only concerns with taking out the node from the tree, but also requires attention on merging the sub-trees below the identified node with its parent node. Let $a$ denote the parent node, and we want to merge the two children of node $a$. Depending on the positions of the two child nodes, i.e., $left(a)$ and $right(a)$, the merging operations can be categorized into three cases: 
\begin{description}
\item[(i) \texttt{MergeLeaves}](Algorithm \ref{alg: merging leaves}): When both children are leaf nodes, we merge the two by removing them from the tree, and by passing their identifier to their parents, i.e., $id(a, \mathcal{D}) = id(left(a), \mathcal{D})\cup id(right(a), \mathcal{D})$. See Figure (\ref{fig: MergeLeafChildren}) for a graphical illustration. 
\item[(ii) \texttt{MergeLeafAndNonLeaf}](Algorithm \ref{alg: merging leaf and non-leaf}): When one child of $a$ is leaf, denoted by $c_2$, and the other child is not a leaf, denoted by $c_1$, we connect the sub-tree below $c_1$ with the parent of $c_1$. In this case, the identifier of $c_2$ gets passed to the leaf nodes below $c_1$. See Figure (\ref{fig: MergeLeafAndNonLeafChildren}). 


\item[(iii) \texttt{MergeNonLeaves}](Algorithm \ref{alg: non-leaf nodes}): When both children are not leaves, denoted by $c_1$ and $c_2$, we first connect the sub-tree below $c_1$ with the parent of $c_1$. Then we connect the sub-tree below $c_2$ with each leaf below $c_1$ (node $B$, $C$ and $right(c_1)$ in Figure (\ref{fig: MergeNonLeafChildren})) while passing the identifiers to the leaves in the merged tree. This merging can cause some empty cohort due to conflicting constraints and we prune the tree by removing nodes containing empty cohorts. See Fig (\ref{fig: MergeNonLeafChildren}) for an example. 

\end{description}

Note that whenever we remove a leaf node $\ell$ from the current tree $\mathcal{D}_{current}$ in \texttt{MergeLeaves} and \texttt{MergeLeafAndNonLeaf} or transform a leaf node $\ell$ to non-leaf node in $\mathcal{D}_{current}$ in \texttt{MergeNonLeaves}, we make to pass the identifiers to all leaf nodes $\ell'$ in resulting tree $\mathcal{D}_{new}$ such that $\Pi_{\mathbf{X}\setminus \mathbf{X}'}(\ell',\mathcal{D}_{new}) \subseteq \Pi_{\mathbf{X}\setminus \mathbf{X}'}(\ell,\mathcal{D}_{current})$. This facilitates the retrieval of $S_{\ell}$ defined in \eqref{eq: super-cohot} from the output $\mathcal{D}_{out}$ of Algorithm \ref{alg: remove nodes} as $S_{\ell} = \{ \ell' \in \mathcal{D}_{in} : id(\ell', \mathcal{D}_{in}) \in id(\ell, \mathcal{D}_{out})\}$.

  \begin{algorithm}[!htb]
    \caption{MergeLeaves($\mathcal{D}$, $a$)}
    \small
    \label{alg: merging leaves}
    \begin{algorithmic}[1]
        \REQUIRE a decision tree $\mathcal{D}$, a node $a$ such that both children of $a$ are leaf nodes
        \STATE Set $id(a, \mathcal{D}) = id(left(a), \mathcal{D})\cup id(right(a), \mathcal{D})$;
        \STATE Set $left(a) = right(a) = \emptyset$;
        \STATE Return $\mathcal{D}$;
    \end{algorithmic}
\end{algorithm}  

\begin{algorithm}[!htb]
    \caption{MergeLeafAndNonLeaf($\mathcal{D}$, $a$)}
    \label{alg: merging leaf and non-leaf}
    \begin{algorithmic}[1]
        \REQUIRE a decision tree $\mathcal{D}$, a node $a$ such that one child of $a$ is a leaf and the other child is not a leaf
        \STATE Let $c_1$ be the non-leaf child of $a$ and let $c_2$ be the leaf child;
        \STATE Set $left(a, \mathcal{D}) = left(c_1)$ and $right(a) = right(c_1)$;
        \FOR {each visited node $a_1$ in a breadth-first traversal in the sub-tree below $a$}
            \IF {$a_1$ is a leaf node}
                \STATE Set $id(a_1, \mathcal{D}) = id(a_1, \mathcal{D})\cup id(c_2, \mathcal{D})$;
            \ENDIF
        \ENDFOR
        \STATE Return $\mathcal{D}$;
    \end{algorithmic}
\end{algorithm}

\begin{algorithm}[H]
    \caption{MergeNonLeaves($\mathcal{D}$, $a$)}
    \label{alg: non-leaf nodes}
    \begin{algorithmic}[1]
        \REQUIRE a decision tree $\mathcal{D}$, a node $a$ such that both children of $a$ are non-leaf nodes
        \STATE Let $c_1 = left(a)$ and $c_2 = right(a)$;
        \STATE Set $left(a) = left(c_1)$ and $right(a) = right(c_1)$;
        \FOR {each visited node $a_1$ in a breadth first traversal in the sub-tree below $a$}
            \IF {$a_1$ is a leaf node}
                \STATE Set $left(a_1, \mathcal{D}) = left(c_2, \mathcal{D})$ and $right(a_1, \mathcal{D}) = right(c_2, \mathcal{D})$;
                \FOR {each visited node $a_2$ in a breadth first traversal in the sub-tree starting from $a_1$}
                    \IF {$a_2$ is a leaf node}
                        \STATE $id(a_2, \mathcal{D}) = id(a_1, \mathcal{D}) \cup id(a_2, \mathcal{D})$;
                    \ELSE
                        \IF {$\Pi_{\mathbf{X}}(c, \mathcal{D})  = \emptyset$ for a child of $a_2$}
                            \STATE Let $c'$ be the other child;
                            \STATE Remove $c$ and $c'$ by setting $left(a_2) = left(c')$ and $right(a_2) = right(c')$ and
                            \IF{$c'$ is a leaf}
                                \STATE $id(a_2, \mathcal{D}) = id(c', \mathcal{D})$;
                            \ENDIF
                        \ENDIF
                    \ENDIF
                \ENDFOR
            \ENDIF
        \ENDFOR
        \STATE Return $\mathcal{D}$;
    \end{algorithmic}
\end{algorithm}

\subsection{The Complete Algorithm}\label{subsection: GCT algo}
We formally present the GCT algorithm for identifying heterogeneous cohorts and estimating the treatment effects of those cohorts in Algorithm \ref{alg: generalized causal tree}. The algorithm is 
based on the implementation-friendly objective function defined in Section \ref{subsection: implementation-friendly modification} and user-friendly output given in Section \ref{subsection: user-friendly output}.

\begin{algorithm}[!ht]
    \caption{Generalized Causal Tree}
    \label{alg: generalized causal tree}
    \begin{algorithmic}[1]
        \STATE Learn a causal tree $\mathcal{C}_{\mathbf{X}, Z}$ with treatment W, heterogeneity factors $(\mathbf{X}, Z)$ and response $Y$;
        \STATE Let $\Lambda_1,\ldots,\Lambda_M$ be the leaf-cohorts of $\mathcal{C}_{\mathbf{X},Z}$ corresponding to leaf nodes $\ell_1, \ldots, \ell_M$;
        \STATE Let $\hat{\tau}(\Lambda_i) := \hat{E}[Y(W = 1) | (\mathbf{X}, Z) \in \Lambda_i] - \hat{E}[Y(W = 0) | (\mathbf{X}, Z) \in \Lambda_i]$ for $i = 1, \ldots, M$;
        \STATE Remove all nodes involving $Z$ from $\mathcal{C}_{\mathbf{X}, Z}$ using Algorithm \ref{alg: remove nodes} with $\mathcal{D} = \mathcal{C}_{\mathbf{X}, Z}$ and $\mathbf{X}' = \{Z\}$ and denote the output by $\mathcal{C}_{\mathbf{X}}$ with leaf-cohorts $\Pi_1, \ldots, \Pi_{K_{\mathcal{X}}}$ corresponding to leaf nodes $\ell_{\mathbf{X}, 1}, \ldots, \ell_{\mathbf{X}, K_{\mathcal{X}}}$;
        \STATE Remove all nodes involving $\mathbf{X}$ from $\mathcal{C}_{\mathbf{X}, Z}$ using Algorithm \ref{alg: remove nodes} with $\mathcal{D} = \mathcal{C}_{\mathbf{X}, Z}$ and $\mathbf{X}' = \mathbf{X}$ and denote the output by $\mathcal{C}_{Z}$ with leaf-cohorts $\{\Gamma_1, \ldots, \Gamma_{K_{\mathcal{Z}}} \}$ corresponding to leaf nodes $\ell_{Z, 1}, \ldots, \ell_{Z, K_{\mathcal{Z}}}$;
        \FOR {each $(\Pi_i, \Gamma_j)$}
            \STATE Find the unique leaf node $\ell$ in $\mathcal{C}_{\mathbf{X}, Z}$ corresponding to the id $id(\ell_{\mathbf{X}, i}, \mathcal{C}_{\mathbf{X}}) \cap id(\ell_{Z, j}, \mathcal{C}_{Z})$; 
            \STATE Set $\hat{\tau}(\Pi_i, \Gamma_j) = \hat{\tau}(\Lambda_{\ell})$;
        \ENDFOR
    \end{algorithmic}
\end{algorithm}

The output of the algorithm is a partition $\{\Pi_1, \ldots, \Pi_{K_{\mathcal{X}}}\}$ of $\mathcal{X}$, a partition $\{\Gamma_1, \ldots, \Gamma_{K_{\mathcal{Z}}} \}$ of $\mathcal{Z}$, and the treatment effects $\hat{\tau}(\Pi_i, \Gamma_j)$ for $i = 1, \ldots, K_{\mathcal{X}}$ and $j = 1,\ldots, K_{\mathcal{Z}}$. Given these estimates, for any $x \in \mathcal{X}$ and $z \in \mathcal{Z}$, we can estimate the treatment effect as $\hat{\tau}(\Pi_i, \Gamma_j)$ by identifying the unique cohorts $\Pi_i$ and $\Gamma_j$ such that $x \in \Pi_i$ and $z \in \Gamma_j$. A step-by-step demonstration of Algorithm \ref{alg: generalized causal tree} 
is provided in Appendix A.

\vspace{2ex}
{\bf Theoretical Properties:} Our algorithm is designed in a way such that all statistical properties of the single treatment causal tree based treatment effect estimators hold for our GCT based treatment effect estimators. This follows from the following identifiability result where we show that the causal tree algorithm based on the treatment indicator $W$ and the heterogeneity factors $(X, \mathbf{Z})$ is indeed estimating $\tau(z, \mathbf{x})$. The proof is given in Appendix B. 
\begin{theorem}\label{theorem: identifiability}
Let $\tau(z, \mathbf{x})$ be the conditional average treatment effect of $T = z$ given $\mathbf{X} = x$. Then  $\tau(z, \mathbf{x}) = E[Y(W=1) \mid \mathbf{X} = \mathbf{x}, Z = z] - E[Y(W=0) \mid \mathbf{X} = \mathbf{x}, Z = z]$ under the assumptions given in Section \ref{section: intro}.
\end{theorem}


{\bf Scalability:} Since we are just adding one extra variable $Z$ to the feature set $\mathbf{X}$, the impact on the time-complexity of the original causal-tree algorithm (or any decision tree based method) is expected to be minimal. It is easy to show that the worst-case time-complexity of creating the used-friendly output from a given causal tree with $m$ nodes is $\mathcal{O}(m^2)$.

\section{Experiments} \label{section: simulation}



We conduct simulation studies to demonstrate the efficacy of
GCT for handling multiple discrete or continuous treatments.
Here, we consider (i) continuous treatment: $Z$ following a uniform distribution on $\mathcal{Z} = (0,1]$, (ii) ordinal treatment: $Z$ following a uniform distribution on the ordered set $\mathcal{Z} = \{1,2,3,4,5,6\}$, and (iii) categorical treatment: $Z$ follows a uniform distribution on the set $\mathcal{Z} = \{a,b,c,d\}$.
Note that the GCT algorithm treats in the ordinal and categorical treatment cases slightly differently. In the ordinal case, GCT anticipates the function $\tau(\mathbf{x}, t)$ to be a smooth function of $t$ for each $\mathbf{x}$ and considers splitting the treatment values $\{1,2,\ldots,M\}$ into two groups (while building the tree) based on $M-1$ choices (i.e., $\{1,\ldots, i\}$ and  $\{i+1,\ldots,M\}$ for $i$ = $1,\ldots,M-1$) instead of all $2^M$ choices (e.g., the split $\{1, M\}$ and $\{2,\ldots,M-1\}$ is not considered in the ordinal case). 

For each setup, we generate a sample of $N = 2000$ individuals with heterogeneity features $\mathbf{X} = (X_{1}, X_{2})$, where $X_1$ and $X_2$ are independently generated from a $Uniform(0,1)$ distribution. The sample is randomly splitted into two halves: one for training (denoted by $\{\mathbf{X}^{tr}\}$) and one for test  (denoted by $\{\mathbf{X}^{te}\}$). On the training data, we independently generate the treatments $T_i^{tr}$ for each individual $i$. Conditioning on $(\mathbf{X}_i^{tr}, T_i^{tr})$, the outcome $Y_i^{tr}$ is generated according to $Y_i^{tr} = f(\mathbf{X}_i^{tr}, T_i^{tr}) + \epsilon_i^{tr}$, where $\epsilon_i^{tr}$'s are i.i.d.\ $\mathcal{N}(0, 1)$ random variables and $f(\cdot, \cdot)$ is defined below. From the training data, we learn $\mathbf{X}$-cohorts $(\Pi_1, \cdots, \Pi_K)$ along with the within-cohort optimal treatment set $\Gamma_{j_r}$ for each $\Pi_r$. Then, on the test data, for each $\mathbf{X}_i^{te}$, we determine the cohort $\Pi_r$ such that $\mathbf{X}_i^{te} \in \Pi_r$ and generate $T_i^{te}$ by randomly choosing a value from the uniform distribution on $\Gamma_{j_r}$. Conditioning on $(\mathbf{X}_i^{te}, T_i^{te})$, the outcome $Y_i^{te}$ is generated according to $Y_i^{te} = f(\mathbf{X}_i^{te}, T_i^{te}) + \epsilon_i^{te}$, where $\epsilon_i^{te}$'s are i.i.d.\ $\mathcal{N}(0, 1)$ random variables. 
We define $f(\cdot,\cdot)$ through the function $\eta(x_1, x_2) = -2 +  4/([1+\exp(-12(x_1 - 0.2))][1+\exp(-12(x_2 - 0.2))])$.
of heterogeneous features $\mathbf{X}_i = (X_{1i}, X_{2i})$. Figure \ref{fig:heatmap-2:3} in the appendix provides a graphical illustration of the values of $\eta(x_1,x_2)$ for $(x_1,x_2)$ within a unit square. Blue and red depict the regions in which $\eta(x_1, x_2)$ takes positive and negative values, respectively. The heterogeneous nature of how the function $\eta(x_1, x_2)$ responds to features makes it the building block for designing our experiments. Besides $\eta(x_1,x_2)$, we include $\eta(x_1,1-x_2)$ to infuse more heterogeneity to the data. We define different $f(\cdot,\cdot)$ functions corresponding to the different types of treatments such as continuous, ordinal, categorical.
In Appendix C, 
we provide discussions on the choice and the intuition behind our simulation designs. 

\begin{figure}[!htb]
    \centering
    \includegraphics[width=0.48\textwidth]{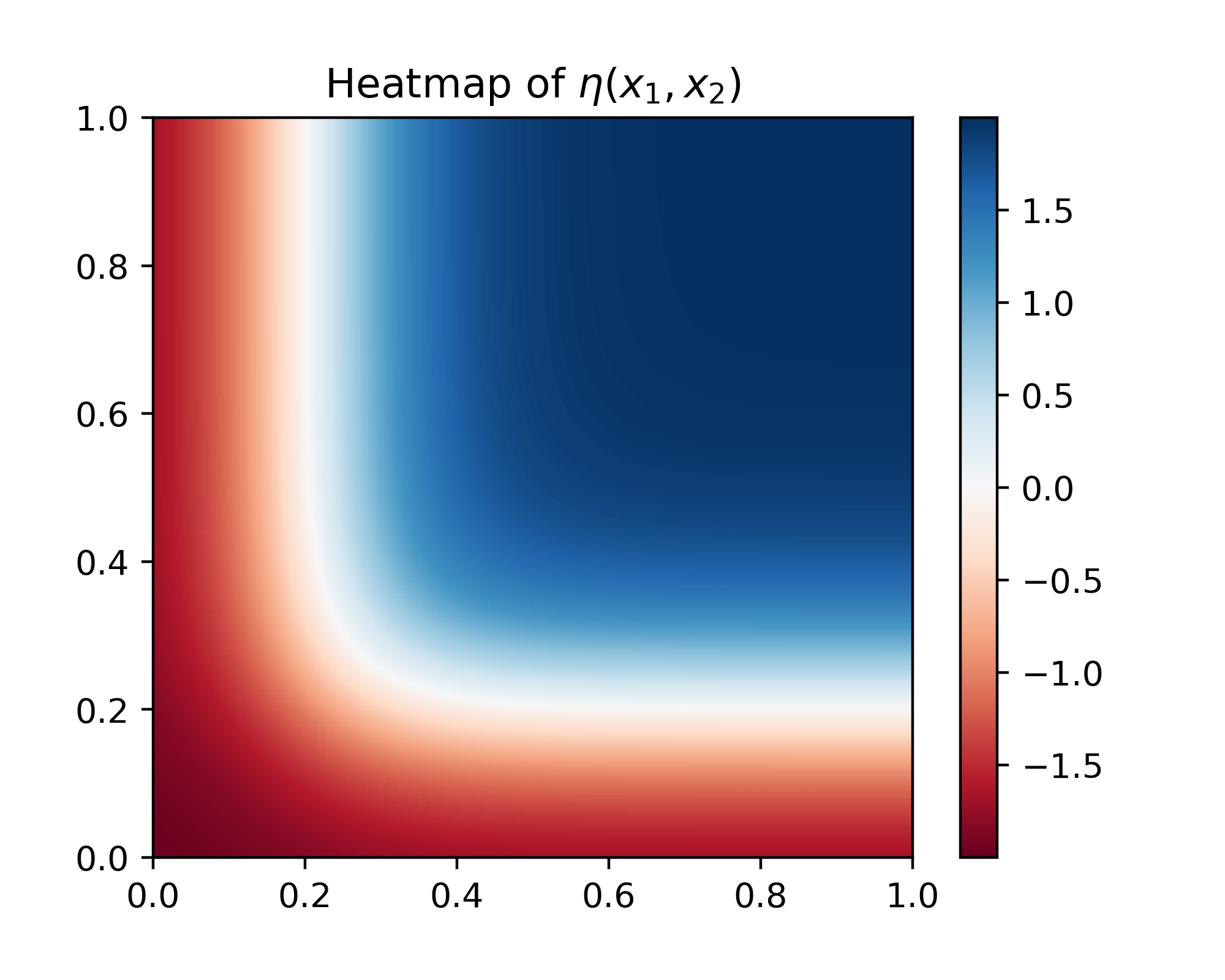}
    \includegraphics[width=0.48\textwidth]{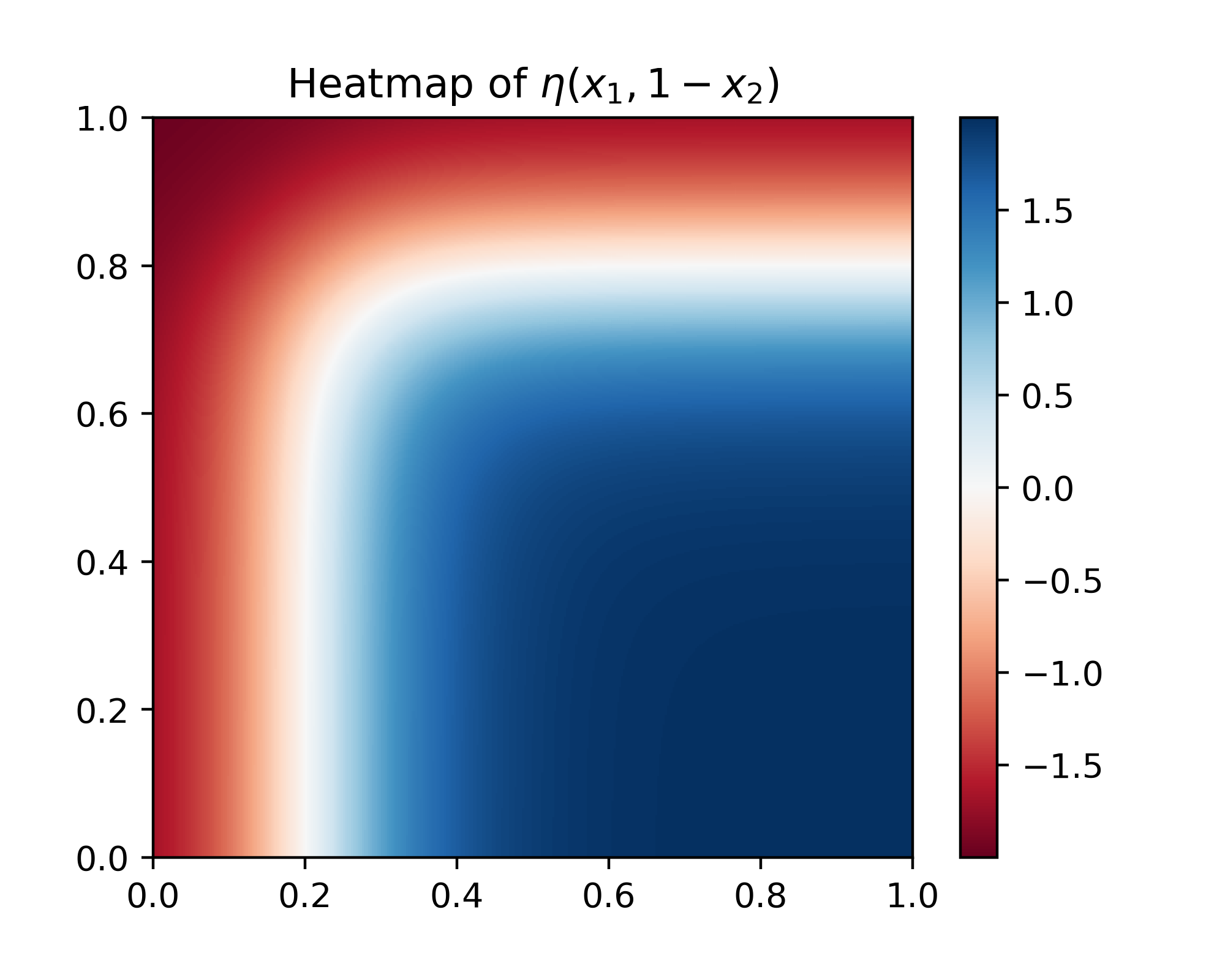}
    \caption{\small Heatmaps of $\eta(x_1, x_2)$ and $\eta(x_1, 1-x_2)$.}
    \label{fig:heatmap-2:3}
\end{figure}

We compare GCT\footnote{The code is available through \href{https://github.com/xiufanyu/GCT}{\textcolor{blue}{Github}}.} with existing decision tree based uplift modeling, including the original causal tree \citet{AtheyImbens16} and multi-treatment uplift models of \citet{RadcliffeSurry11,zhao2017uplift} based on various split criteria including the KL divergence (KL), Euclidean distance (ED), Chi-Square (CHI), and contextual treatment selection (CTS)\footnote{CT is implemented using ``causalTree'' R package, and the latter four methods are implemented using ``CausalML'' Python package \citep{chen2020causalml}.}. Since existing methods are not defined for continuous treatments, we discretize the continuous treatment into $4$ equally sized bins for these methods and use the mid-point of each bin to represent the treatment level for that bin. Recall that the original causal tree algorithm can only work with a single treatment. Therefore, when learning the heterogeneous cohorts using CT, we have to ignore the treatment values $Z$ but only use the treatment indicator $W$. The resultant estimated causal effects for each cohort are computed on a treatment-vs-control basis regardless of treatment values. To assign the optimal treatment, we randomly choose a value from a uniform distribution on $\mathcal{Z}$ whenever the treatment effect is positive. We refer to this variation of causal tree by CT-B (where B stands for ``binary''). We consider another variation, called CT-M (M for ``multiple''), where we estimate causal effects for every discrete (or discretized) treatment level separately within each cohort after applying the causal tree based on $W$, and we assign optimal treatments based on these estimates.



\begin{figure}[!htb]
\centering
\includegraphics[width = 0.8 \linewidth]{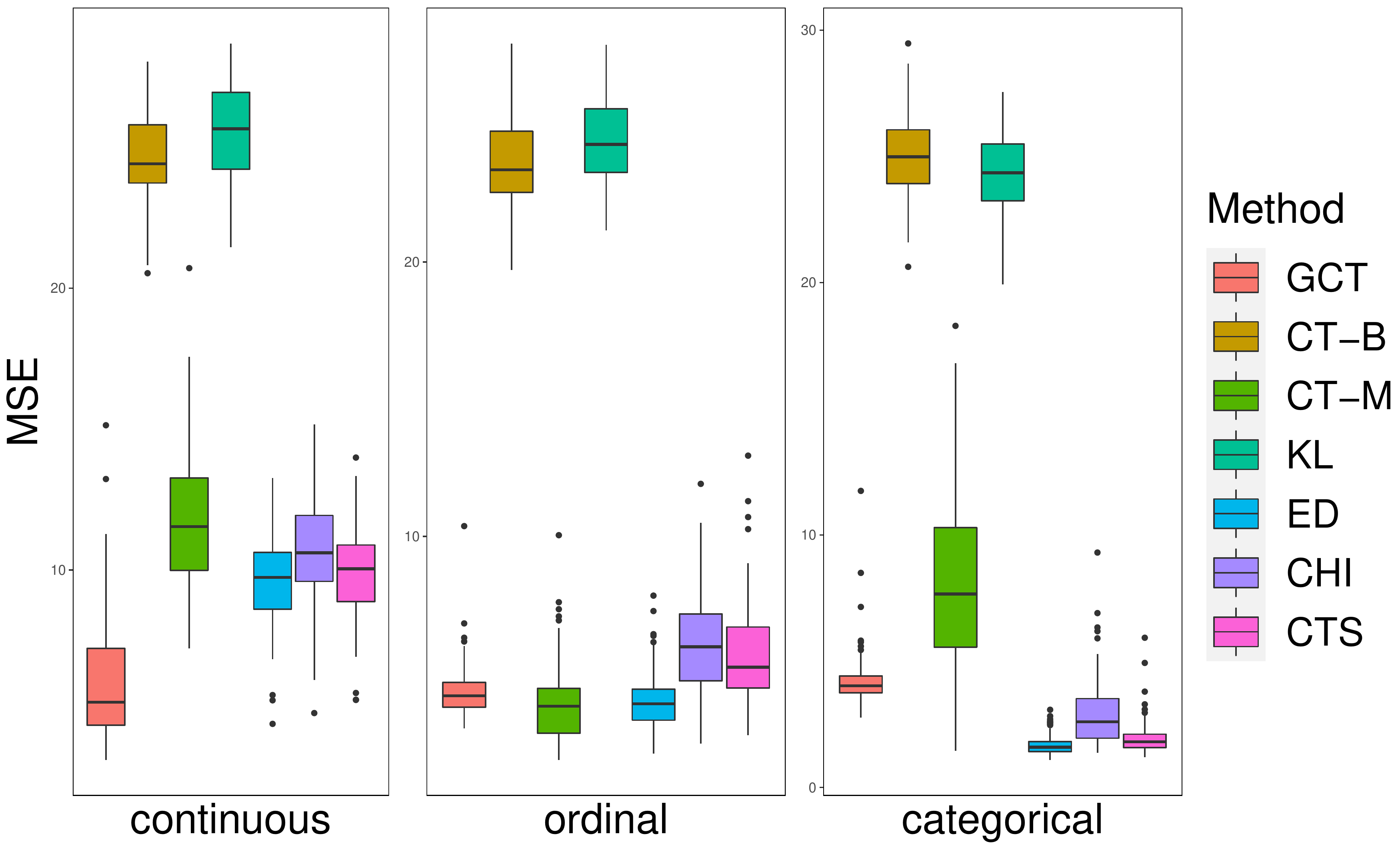}
\captionsetup{justification=centering}
\caption{\small Box-plots of the MSE of estimated treatment effects \\ based on 100 iterations.}\label{Fig:MSE}
\end{figure}

The algorithm performance is evaluated by two metrics: (1) the mean square error (MSE) of estimated causal effects, 
and (2) the average outcome with the optimal treatment allocated to the testing data. The former measures the accuracy of an algorithm in estimating heterogeneous treatment effects, while the latter measures uplifting quality based on the optimal treatment allocation according to different algorithms. Figure \ref{Fig:MSE} shows box-plots of MSE values. Results 
are summarized in Table \ref{tab:AverageOutcome}. Each metric is averaged over 100 replications. The standard errors are reported in the parentheses.

From Figure \ref{Fig:MSE}, we can see that GCT achieves the smallest MSE value in the continuous setting and yields comparable MSE with ED, CHI, and CTS in the ordinal and categorical settings. More importantly,  Table \ref{tab:AverageOutcome} shows that we can achieve the highest average outcome by determining the optimal treatment allocations based on the output of GCT. The negative outcome obtained by CT-B in the categorical setting further implies that neglecting the existence of heterogeneous effects among various treatments may lead to unfavorable results.

\begin{table}[!htb]
\centering
\captionsetup{justification=centering}
    \caption{Average outcome with optimal treatment allocations (a higher value implies a better performance).}\label{tab:AverageOutcome}
    \begin{tabular}{c|ccccccc} 
    \toprule
    & \multicolumn{3}{c}{Treatment Type} \\
    \cline{2-4}
    Method & Continuous & Ordinal & Categorical \\ 
    \hline
    GCT  & \textbf{5.744} (0.051) &  \textbf{5.321} (0.048) & \textbf{4.480} (0.045)\\ 
    CT-B &  0.188 (0.013)  &  0.581 (0.017)  & -0.008 (0.012) \\
    CT-M &  4.880 (0.092)  &  3.904 (0.033)  &  3.618 (0.057) \\   
    KL   &  2.519 (0.033)  &  2.519 (0.033)  & 2.502 (0.039) \\ 
    ED   &  0.548 (0.036)  &  0.417 (0.041)  & 0.466 (0.033) \\
    CHI  &  0.698 (0.038)  &  0.843 (0.050)  & 0.626 (0.037) \\
    CTS  &  0.598 (0.035)  &  0.818 (0.050)  & 0.510 (0.035) \\
    \bottomrule
    \end{tabular}
\end{table}

\begin{table}[!hbt]
    \centering
    \caption{Expected response $\bar{R} = \frac{1}{N}\sum_{i=1}^N R_i$.}
    \label{tab:realdata2}
    \begin{tabular}{cccccccc} 
    \toprule
    Dataset & GCT & CT-B & CT-M & KL & ED & CHI & CTS\\ 
    \midrule
   {Bladder} & 0.81 & 0.76 & 0.79 & \textbf{0.82} & 0.81 & 0.82 & 0.81 \\ 
   {AOD} & \textbf{0.62} & 0.58 & 0.60 & 0.61 & 0.60 & 0.56 & 0.59 \\
    \bottomrule
    \end{tabular}
\end{table}

\section{Real Data Analysis} \label{section: realdata}

In a recent benchmarking study of multi-treatment uplift modeling methods \citep{olaya2020survey}, the authors compared tree-based methods with other (less-interpretable) methods to conclude that no approach is strictly superior to the others. We used two datasets from the same benchmarking setup to compare our method with the tree-based methods and we refer to \cite{olaya2020survey} for a comparison with the other multi-treatment uplift modeling methods.
In particular, we apply the proposed GCT as well as CT-B, CT-M, and the four tree-based methods KL, ED, CHI, and CTS to two real datasets to evaluate their uplifting effects. 
One is the \emph{Bladder} dataset on the recurrence of bladder cancer with weak heterogeneous effect, available in the R package \texttt{survival} \citep{survival-package}.
The other is the \emph{AOD} dataset on alcohol and drug usage with a strong heterogeneous effect, available in the R package \texttt{twang} \citep{twang-package}. 
There are 2 treatment groups and 1 control group in both datasets and the outcome for Bladder is binary while the outcome for AOD is continuous. Following \citet{olaya2020survey}, we convert the continuous outcome to a binary outcome in AOD for a fair comparison. 
Unlike the simulation studies, it is impossible to re-run the experiments based on the optimal treatment determined by our algorithm. Instead, we follow the proposal of \citet{zhao2017uplift} for quantifying the performance of an algorithm. The individual expected response is defined as $R_i = \sum\nolimits_{k=1}^K \frac{y_i}{P_{k}} \mathds{1}_{\{ v(\mathbf{x}_i) = k\}} \mathds{1}_{\{ T_i = k\}},$
where $P_{k}$ is the empirical probability of an individual receiving treatment $k$, and $v(\mathbf{x}_i)$ is the optimal treatment assigned by the uplift model based on its heterogeneous features $\mathbf{x}_i$.
The expected response of an uplift model is then defined as 
$\bar{R} = \frac{1}{N}\sum_{i=1}^N R_i,$
which is an unbiased estimator of $E[y_i|T=v(\mathbf{x}_i)]$.
Following \cite{olaya2020survey}, we first apply propensity scoring matching to debias the non-experimental data. Then we fit models using 5-fold cross-validation, and the results from each round are averaged to obtain the overall performance.




Table \ref{tab:realdata2} summarizes the average expected responses for the 7 methods. Notice that results of KL and ED are not exactly the same as but very close to those in \cite{olaya2020survey} due to the randomness in the training-test split. For both datasets, all methods performed similarly. This was expected as the number of treatments is small (i.e., two). 
GCT is expected to be more beneficial with a higher number of treatments (or continuous treatments), as we have shown in simulations.


\section{Discussion}
\label{section: discussion}

We presented a novel generalization of the causal tree algorithm for uplift modeling with multiple discrete or continuous treatments. One of the key features of our generalized causal tree (GCT) algorithm is that it provides a data-driven way of grouping similar treatments together in addition to partitioning the underlying population into disjoint cohorts based on the homogeneity of the treatment effects. We provide significant improvements over the basic version of GCT to make it implementation-friendly and make its output easily adoptable to downstream applications, such as optimal treatment allocation. 
Note that the implementation-friendly modification 
might lead to performance loss by creating unnecessary splits in the control data, which is expected to be negligible when the size of the control data is much larger than the training data. 

We empirically showed that GCT outperforms some extensions of the original causal tree algorithm as well as existing decision tree based multiple treatment uplift modeling methods. Note that our generalization strategy is not limited to the splitting criterion of the causal tree algorithm and hence can be used to extend any decision tree based uplift modeling technique to multiple discrete or continuous treatments. We acknowledge that the decision tree based approaches may not be the best performing methods in certain contexts. However, they might still be preferable because of their interpretability and the computational efficiency in using their output in optimal treatment allocation. Finally, we also note that GCT can be directly used in a continuous treatment version of some recent extensions to uplift modeling, such as treatments with different costs \citep{zhao2019uplift}, and constrained utility optimization \citep{TuEtAl21}.

\bibliographystyle{apalike}
\bibliography{GCT}

\newpage
\appendix

\section*{Appendices}
The appendices provide supplemental details on the algorithms and numerical results to Sections 3-4 of the main context. 
Section \ref{sec: a toy example} uses a toy example to provide a step-by-step demonstration of the implementation of our proposed GCT algorithm. Section \ref{sec:proofs} presents proofs. Section \ref{sec: more on experiments} discusses the simulation designs. 

\section{A Toy Example} \label{sec: a toy example}
\begin{figure}[H]
    \centering
    \includegraphics[width=0.8\textwidth]{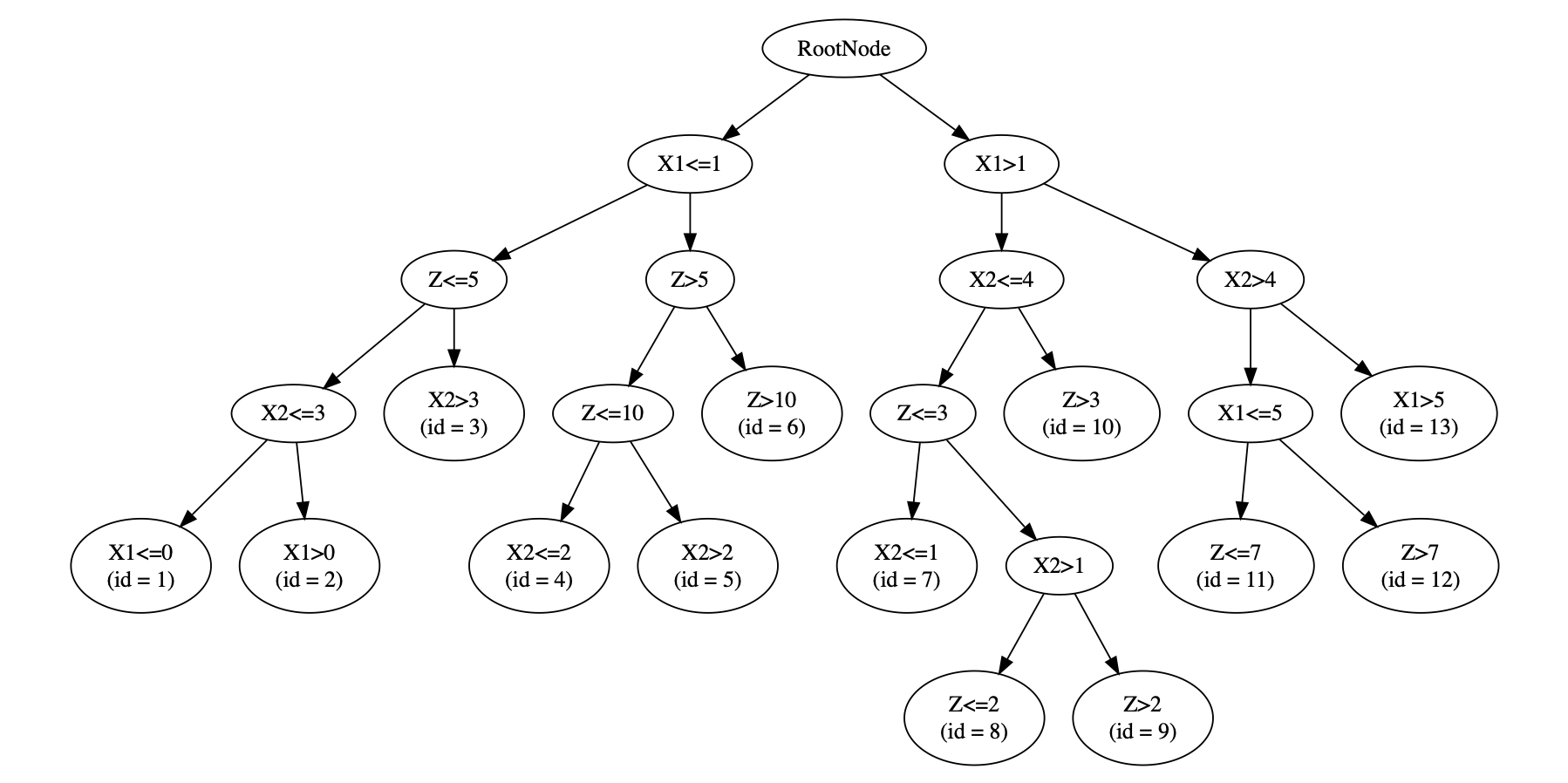}
    \caption{An example of causal tree $\mathcal{C}_{\mathbf{X}, Z}$}
    \label{fig: toyExampleTreeXZ}
\end{figure}

This section presents a toy example of our proposed GCT approach. Figure \ref{fig: toyExampleTreeXZ} plots a causal tree $\mathcal{C}_{\mathbf{X}, Z}$ with respect to heterogeneity factors $\mathbf{X} = (X_1, X_2)$ and treatment values $Z$.
From $\mathcal{C}_{\mathbf{X}, Z}$, we obtain the leaf-cohorts $\Lambda_1,\ldots,\Lambda_M$ with $M = 13$. Each leaf-cohort corresponds to a leaf node in the tree. To be more specific, we summarize the details of all leaf cohorts in Table \ref{tab: leaf-cohorts}.

\begin{table}[H]
    \centering
    \caption{Leaf-cohorts $\Lambda_1,\ldots,\Lambda_M$ from the causal tree $\mathcal{C}_{\mathbf{X}, Z}$} \label{tab: leaf-cohorts}
    \resizebox{\textwidth}{!}{
    \begin{tabular}{cll|cll}
    \hline
    Leaf-cohorts & Leaf nodes & \multicolumn{1}{c|}{Cohort regions} & Leaf-cohorts & Leaf id &\multicolumn{1}{c}{Cohort regions} \\ \hline
    $\Lambda_1$  & $\ell_1$: $\id = 1$ & $\{X_1\leq 0, X_2\leq 3, Z\leq 5\}$     &  $\Lambda_8$    & $\ell_8$: $\id = 8$ & $\{X_1 > 1, 1 < X_2\leq 4, Z \leq 2\}$ \\
    $\Lambda_2$  & $\ell_2$: $\id = 2$ & $\{0<X_1\leq 1, X_2\leq 3, Z\leq 5\}$   &  $\Lambda_9$    & $\ell_9$: $\id = 9$ & $\{X_1 > 1, 1 < X_2\leq 4, 2 < Z \leq 3\}$ \\
    $\Lambda_3$  & $\ell_3$: $\id = 3$ & $\{X_1\leq 1, X_2>3, Z\leq 5\}$         &  $\Lambda_{10}$ & $\ell_{10}$: $\id = 10$ & $\{X_1 > 1, X_2\leq 4, Z > 5\}$ \\
    $\Lambda_4$  & $\ell_4$: $\id = 4$ & $\{X_1\leq 1, X_2\leq 2, 5<Z\leq 10\}$  &  $\Lambda_{11}$ & $\ell_{11}$: $\id = 11$ & $\{1 < X_1 \leq 5, X_2 >4, Z \leq 7\}$ \\
    $\Lambda_5$  & $\ell_5$: $\id = 5$ & $\{X_1\leq 1, X_2>2, 5 < Z\leq 10\}$    &  $\Lambda_{12}$ & $\ell_{12}$: $\id = 12$ & $\{1 < X_1 \leq 5, X_2>4, Z>7\}$ \\
    $\Lambda_6$  & $\ell_6$: $\id = 6$ & $\{X_1\leq 1, X_2\in\mathbb{R}, Z>10\}$ &  $\Lambda_{13}$ & $\ell_{13}$: $\id = 13$ & $\{X_1>5, X_2>4, Z\in \mathbb{R} \}$ \\
    $\Lambda_7$  & $\ell_7$: $\id = 7$ & $\{X_1>1, X_2\leq 1, Z\leq 3\}$         & \\
    \hline
    \end{tabular}
    }
\end{table}

\begin{figure}[!h]
\begin{subfigure}{.5\textwidth}
  \centering
  \includegraphics[width=\linewidth]{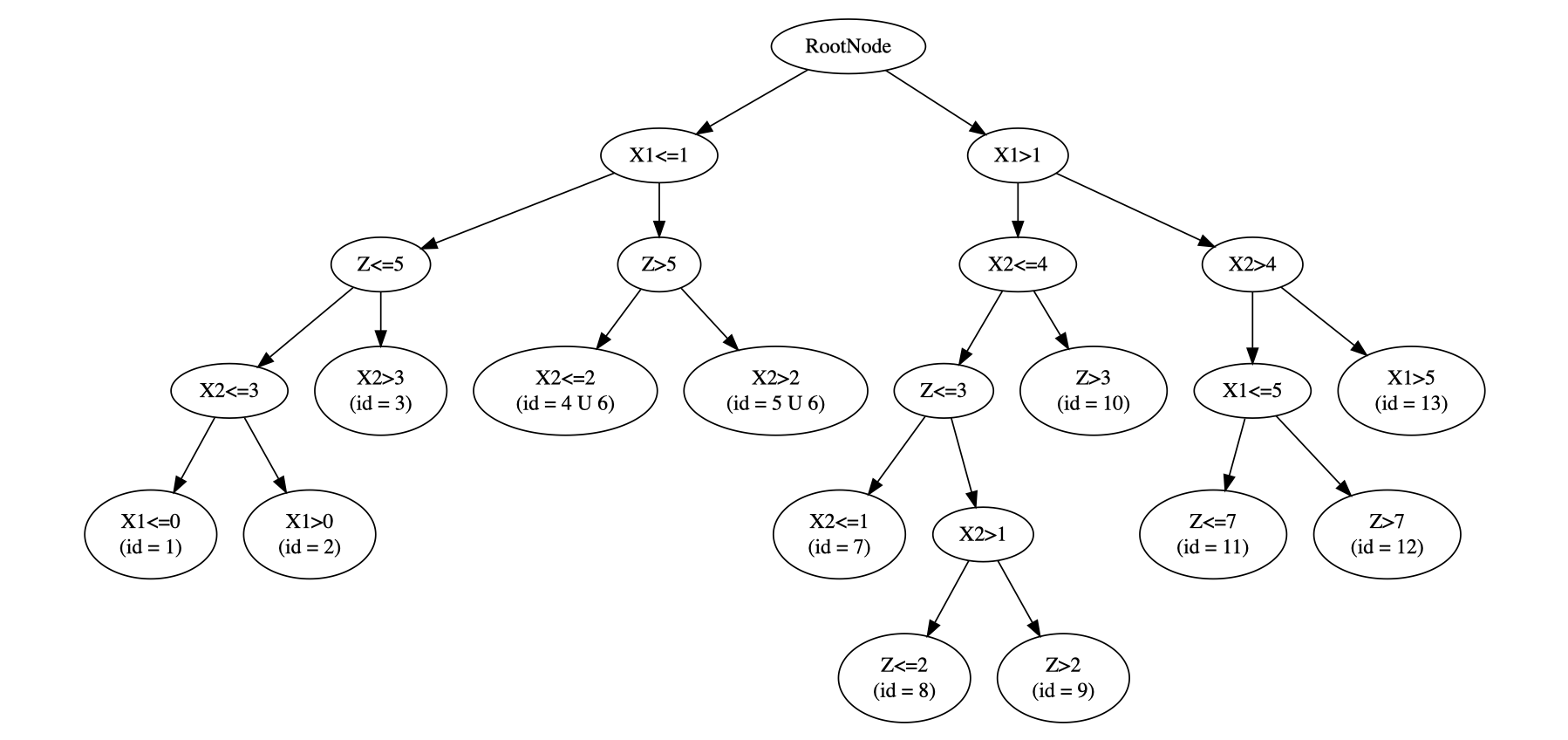}
  \captionsetup{justification=centering}
  \caption{Tree structure after removing the split of \protect\\ $\{Z \leq 10\}$ and $\{Z > 10 \}$.}
  \label{fig: toyExampleRemovalStep1}
\end{subfigure}%
\begin{subfigure}{.5\textwidth}
  \centering
  \includegraphics[width=\linewidth]{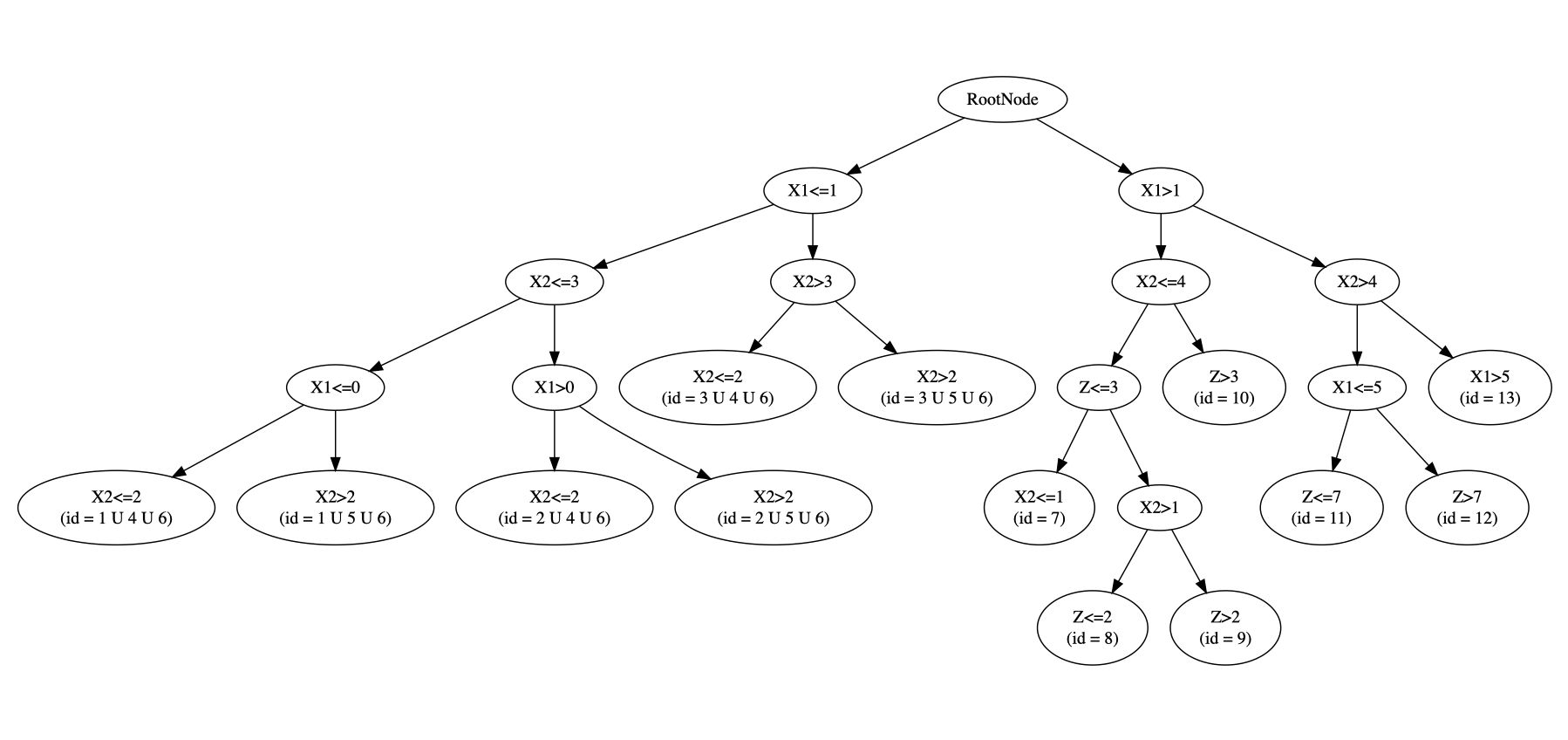}
  \captionsetup{justification=centering}
  \renewcommand\thesubfigure{\alph{subfigure}1}
  \caption{Tree structure after removing the split of \protect\\ $\{Z \leq 5\}$ and $\{Z > 5\}$ .}
  \label{fig: toyExampleRemovalStep2}
\end{subfigure}
\begin{subfigure}{.5\textwidth}
  \centering
  \includegraphics[width=\linewidth]{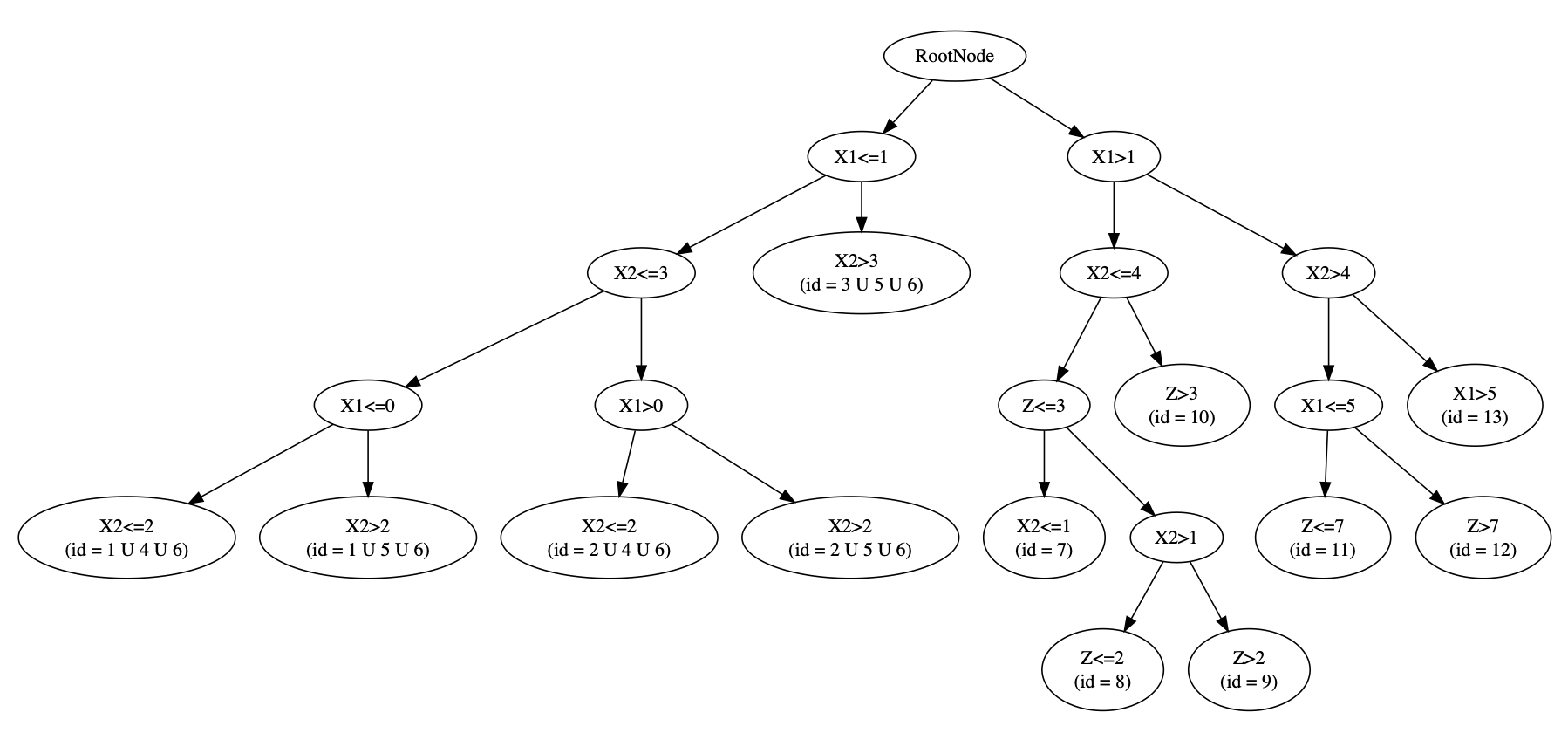}
  \captionsetup{justification=centering}
  \addtocounter{subfigure}{-1}
  \renewcommand\thesubfigure{\alph{subfigure}2}
   \caption{Tree structure after deleting infeasible branches}
  \label{fig: toyExampleRemovalStep2b}
\end{subfigure}%
\begin{subfigure}{.5\textwidth}
  \centering
  \includegraphics[width=\linewidth]{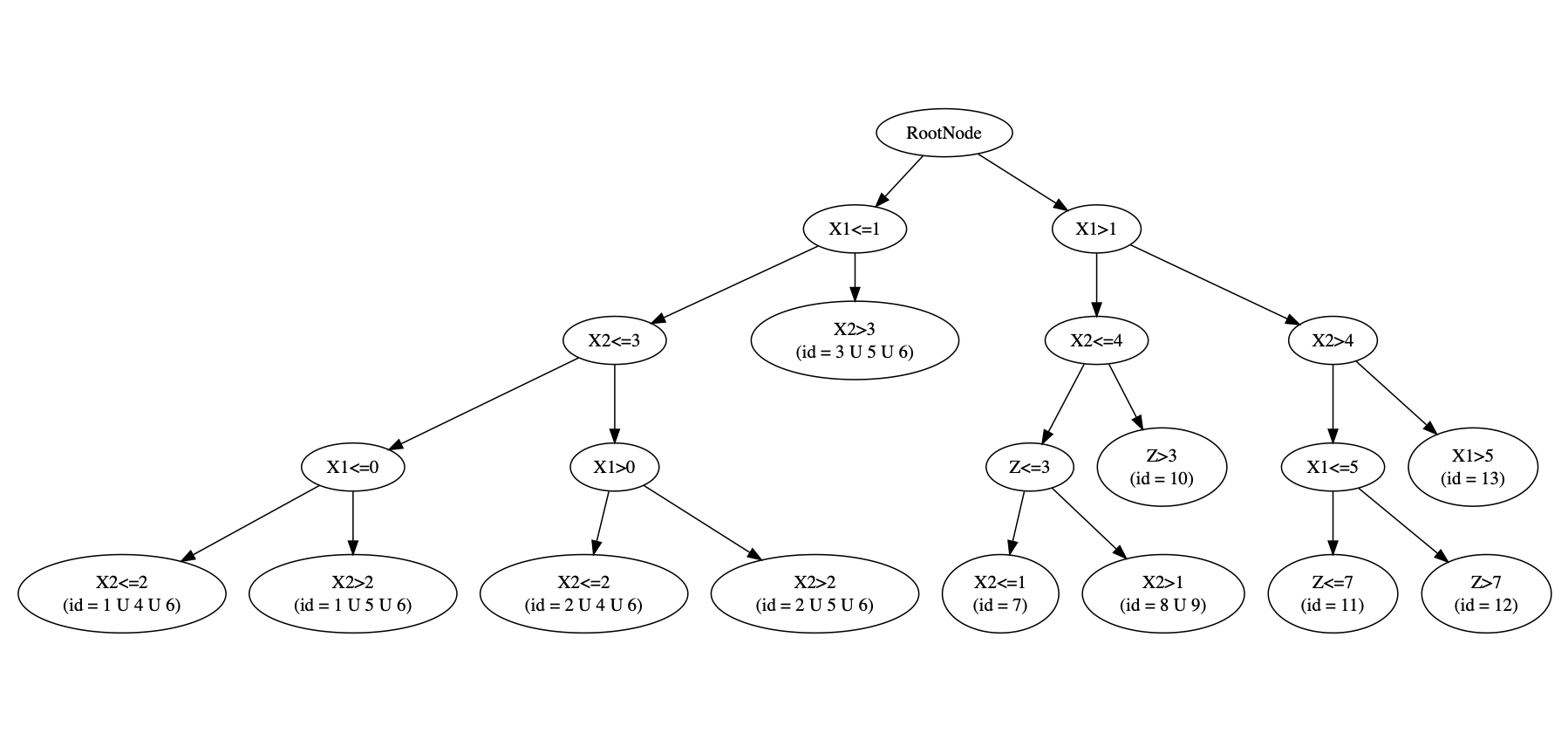}
  \captionsetup{justification=centering}
  \caption{Tree structure after removing the split of \protect\\ $\{Z \leq 2\}$ and $\{Z > 2\}$.}
  
  \label{fig: toyExampleRemovalStep3}
\end{subfigure}
\begin{subfigure}{.5\textwidth}
  \centering
  \includegraphics[width=\linewidth]{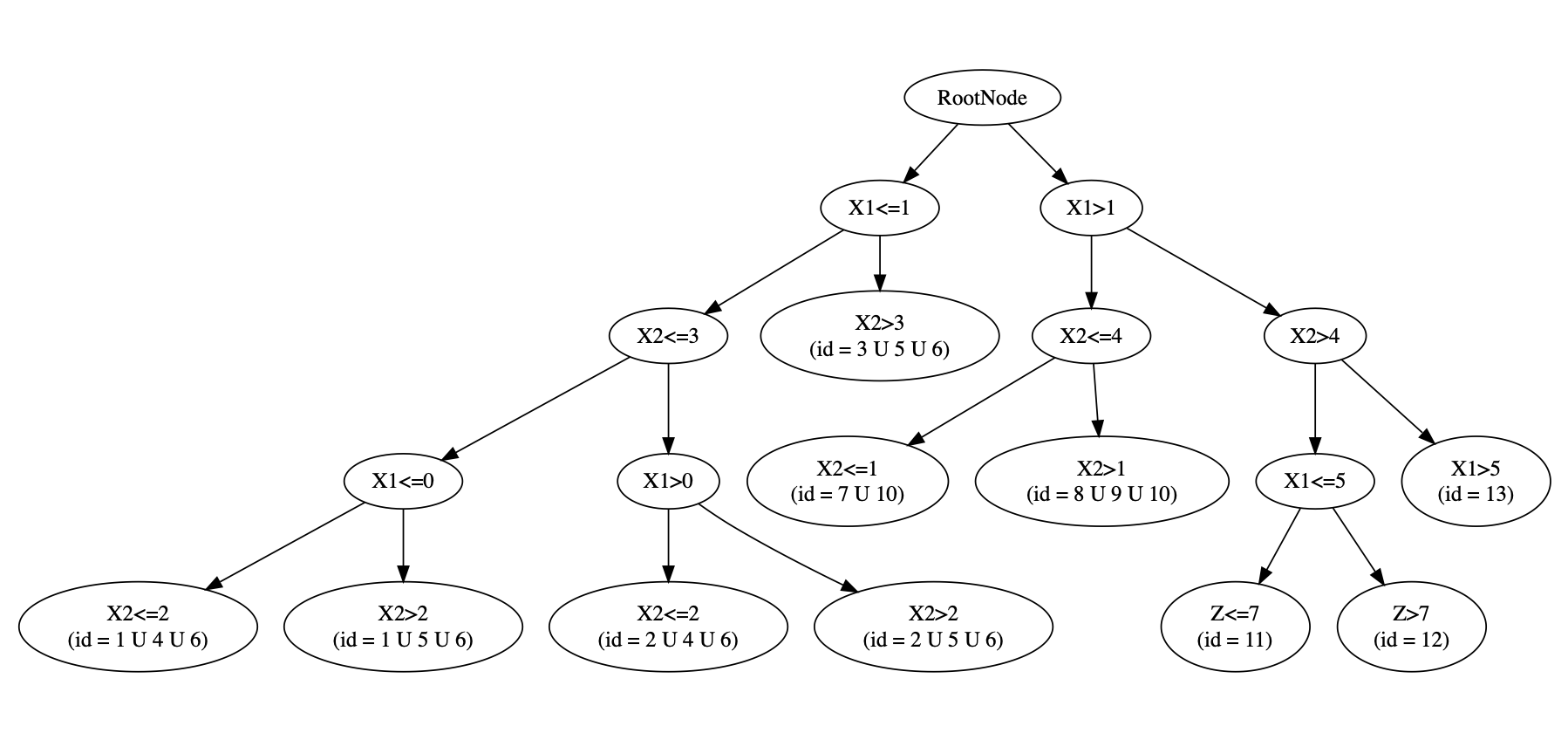}
  \captionsetup{justification=centering}
  \caption{Tree structure after removing the split of \protect\\ $\{Z \leq 3\}$ and $\{Z > 3\}$.}
  \label{fig: toyExampleRemovalStep4}
\end{subfigure}%
\begin{subfigure}{.5\textwidth}
  \centering
  \captionsetup{justification=centering}
  \includegraphics[width=\linewidth]{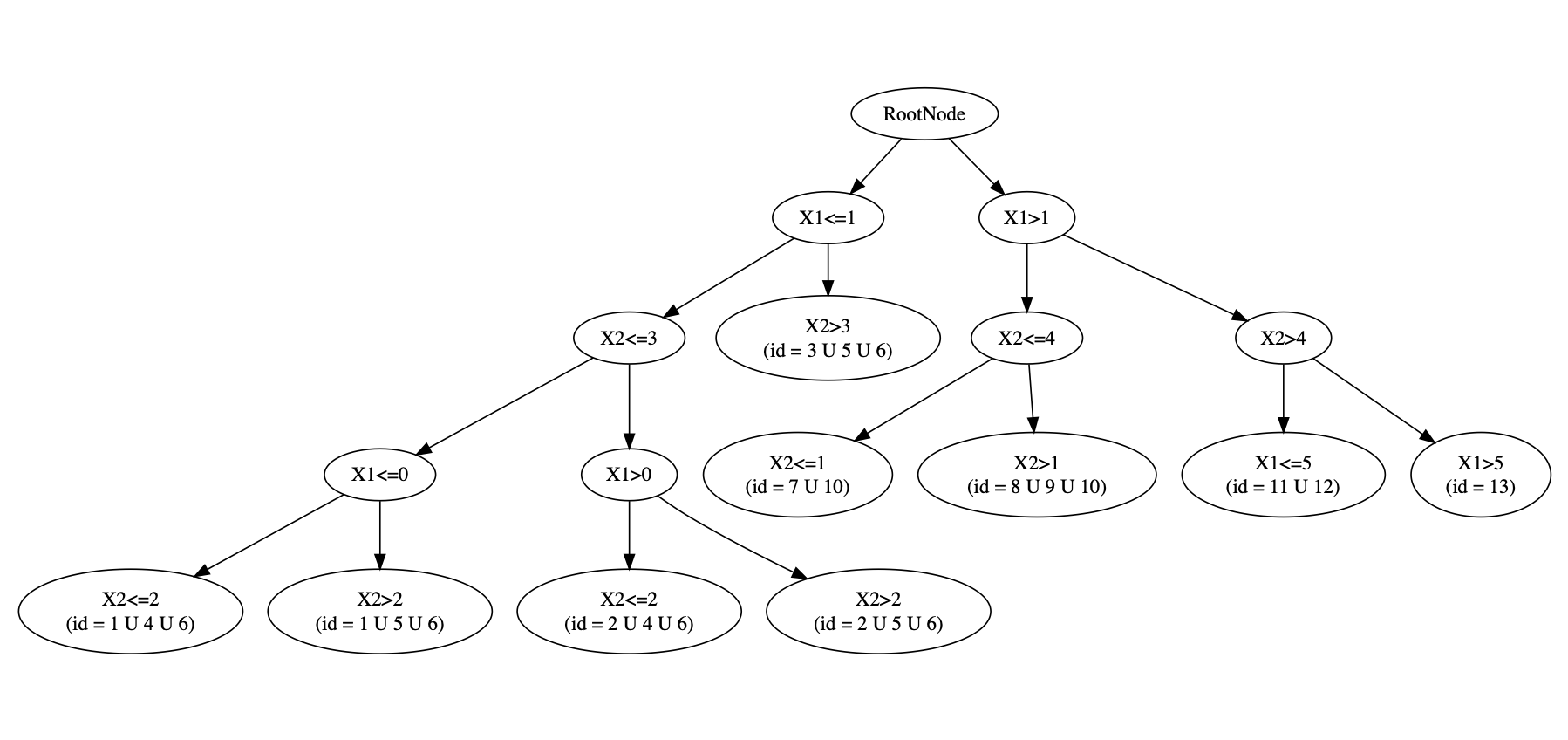}
  \caption{Tree structure after removing the split of \protect\\ $\{Z \leq 7\}$ and $\{Z > 7\}$.}
  \label{fig: toyExampleRemovalStep5}
\end{subfigure}
\captionsetup{justification=centering} \vspace{2ex}
\caption{ Plots of steps to remove $Z$ from $\mathcal{C}_{\mathbf{X}, Z}$ using Algorithm 1. \\ Note: The notation ``id = 1 $\cup$ 2" in the plots means that the samples in this merged node come from leaf nodes $\ell_1$ and $\ell_2$ in reference to the original tree $\mathcal{C}_{\mathbf{X}, Z}$. Please be advised that this does not mean all the samples from the leaf nodes $\ell_1$ and $\ell_2$ are currently in this node with id = 1 $\cup$ 2. }
\label{fig: toyExampleRemovalProcedures}
\end{figure}

Then, we would like to remove all nodes involving $Z$ from $\mathcal{C}_{\mathbf{X},Z}$. In what follows, we present a step-by-step guide to the implementation of Algorithm 1, which is introduced in Section 3.2 of the main context. Figure \ref{fig: toyExampleRemovalProcedures} presents illustrative plots on the structure of the tree after each step. The removal process proceeds in a post-order. 
\begin{itemize}
    \item Step 1: The target node is $a = \{Z>5\}$, removing the split of $\{Z \leq 10\}$ and $\{Z > 10\}$. 
    
    From the graph, we can see that one of $a$'s children (denoted by $c_1$ = {\small $\{Z \leq 10\}$}) is a sub-tree while the other child (denoted by $c_2$ = {\small $\{Z > 10\}$}) is a leaf node. We connect the children of $c_1$ to node $a$ by setting {\small $left(a) = left(c_1)$} and {\small $right(a) = right(c_1)$}, and then merge the node $c_2$ into the leaf nodes of $c_1$, i.e. $\ell_4$ and $\ell_5$ in Figure \ref{fig: toyExampleTreeXZ}. In specific, we modify the ids of $l_4$ and $l_5$ into {\small $id(\ell_4) = id(\ell_4) \cup id (c_2) = 4 \cup 6$} and {\small $id(\ell_5) = id(\ell_5) \cup id (c_2) = 5 \cup 6$}.  Figure (\ref{fig: toyExampleRemovalStep1}) plots the current tree structure after the removal. 
    \medskip
    
    \item Step 2: The target node is $a = \{X_1 \leq 1\}$, removing the split of $\{Z \leq 5\}$ and $\{Z > 5\}$. 
    
    Both of $a$'s children are subtrees. From a high-level point of view, this process consists of two parts: (i) removing the splits associated with $Z$ and merge the branches, followed by (ii) one breadth-first traversal to remove empty branches.
    
    \begin{itemize}
        \item Step 2(i): We connect the children of one subtree ($c_1$ = {\small $\{Z \leq 5\}$}) to $a$'s parent node by setting {\small $left(a) = left(c_1)$} and {\small $right(a) = right(c_1)$}, and then connect the children of another subtree ($c_2$ = {\small $\{Z>5\}$}) to every leaf node of $c_1$, i.e., $\ell_1$, $\ell_2$ and $\ell_3$ in Figure (\ref{fig: toyExampleRemovalStep1}). More precisely, for {\small $a' \in\{\ell_1, \ell_2, \ell_3\}$}, we set {\small $left(a') = left(c_2)$} and {\small $right(a') = right(c_2)$}, in the meantime, we update the ids of all the leaf nodes in $c_2$ (denoted by $a''$) by setting them to be {\small $id(a'') = id(a') \cup id(a'')$}. See Figure (\ref{fig: toyExampleRemovalStep2}) for a graph illustration. 
        
        \item Step 2(ii): We acknowledge that Step 2(i) may result in some infeasible regions because of contradictory conditions. We need one additional step to remove these redundant nodes from the output. To this end, we conduct a level order traversal to remove nodes that correspond to empty regions. For example, in Figure (\ref{fig: toyExampleRemovalStep2}), we observe that the node with {\small $id = 3\cup 4 \cup 6$} (denoted by $c$) corresponds to the path ``{\small $X_1 \leq 1, X_2 > 3, X_2 \leq 2$}", which gives us an empty region as a result of contradictory conditions. Let $a_2$ denote its parent node, i.e. $a_2$ = $\{X_2>3\}$, and let $c'$ denote the sibling of $c$, i.e., $c'$ is the node with {\small $id = 3\cup 5 \cup 6$}. Since $c$ is empty, we remove the split of $c$ and $c'$ from the tree by setting {\small $left(a_2) = left(c')$} and {\small $right(a_2) = right(c')$}. The fact that $c'$ is a leaf node further makes $a_2$ becomes a new leaf node, whose id is set to as $id(a_2) = id(c') = 3\cup 5 \cup 6$. The traversal will help delete infeasible branches so that the output produces non-empty and non-overlapping splits of $\mathbf{X}$. Figure (\ref{fig: toyExampleRemovalStep2b}) presents the final output.
    \end{itemize}
    
    \medskip
    
    \item Step 3: The target node is $a = \{X_2 > 1\}$, removing the split of $\{Z \leq 2\}$ and $\{Z > 2\}$. 
    
    Since both of $a$'s children are leaf nodes, we directly remove the two splits from the tree and convert $a$ into a leaf node, whose id is set to be the union of the previous two leaf nodes, i.e., { \small $id(a) = id(\{Z \leq 2\}) \cup id(\{Z > 2\}) = 8 \cup 9$}. See Figure (\ref{fig: toyExampleRemovalStep3}).
    \medskip
    
    \item Step 4: The target node is $a = \{X_2 \leq 4\}$, removing the split of $\{Z \leq 3\}$ and  $\{Z > 3\}$. 
    
    This is similar to Step 1. One of $a$'s children is a subtree while the other is a leaf node, see Figure (\ref{fig: toyExampleRemovalStep4}) for details. 
    \medskip
    
    \item Step 5: The target node is $a = \{X_1 \leq 5\}$, removing the split of $\{Z \leq 7\}$ and $\{Z > 7\}$. 
    
    This is similar to Step 3 as both of $a$'s children are leaf nodes, see Figure (\ref{fig: toyExampleRemovalStep5}) for the structure after removing this split.
\end{itemize}

At this point, we have finished removing all the nodes associated with $Z$ from the tree. The above output (denoted by $\mathcal{C}_{\mathbf{X}}$) produce a minimal partition with respect to $\mathbf{X}$. The resultant leaf cohorts, $\Pi_1,\ldots,\Pi_K$ with $K = 9$, are summarized in Table \ref{tab: leaf-cohorts-X}.  

\begin{table}[!htb]
    \centering
    \caption{Leaf-cohorts $\Pi_1,\ldots,\Pi_K$ from the causal tree $\mathcal{C}_{\mathbf{X}}$} \label{tab: leaf-cohorts-X}
    \begin{tabular}{cll}
    \hline
    Leaf-cohorts & \multicolumn{1}{c}{Leaf nodes} & \multicolumn{1}{c}{Cohort regions} \\ \hline
    $\Pi_1$  & $a_{\mathbf{X},1}$: $\id = 1\cup 4 \cup 6$ & $\{X_1\leq 0, X_2\leq 2\}$   \\
    $\Pi_2$  & $a_{\mathbf{X},2}$: $\id = 1\cup 5 \cup 6$ & $\{X_1\leq 0, 2<X_2\leq 3\}$ \\
    $\Pi_3$  & $a_{\mathbf{X},3}$: $\id = 2\cup 4 \cup 6$ & $\{0<X_1\leq 1, X_2 \leq 2 \}$        \\
    $\Pi_4$  & $a_{\mathbf{X},4}$: $\id = 2\cup 5 \cup 6$ & $\{0<X_1\leq 1, 2 < X_2\leq 3 \}$  \\
    $\Pi_5$  & $a_{\mathbf{X},5}$: $\id = 3\cup 5 \cup 6$ & $\{X_1\leq 1, X_2>3 \}$    \\
    $\Pi_6$  & $a_{\mathbf{X},6}$: $\id = 7\cup 10$ & $\{X_1>1, X_2\leq 1 \}$ \\
    $\Pi_7$  & $a_{\mathbf{X},7}$: $\id = 8\cup 9 \cup 10$ & $\{X_1>1, 1 < X_2\leq 4\}$         \\
    $\Pi_8$  & $a_{\mathbf{X},8}$: $\id = 11\cup 12$ & $\{1<X_1\leq 5, X_2>4 \}$         \\
    $\Pi_9$  & $a_{\mathbf{X},9}$: $\id = 13$ & $\{X_1>5, X_2>4\}$         \\
    \hline
    \end{tabular}
\end{table}

Similarly, by removing all nodes involving $\mathbf{X}$ from $\mathcal{C}_{\mathbf{X}, Z}$, we obtain a tree consisting of nodes related to $Z$, denoted by $\mathcal{C}_{Z}$. A a minimal partition of $range(Z)$ can be extracted from the leaf-cohorts $\Gamma_1,\ldots, \Gamma_L$. The results are summarized in Table \ref{tab: leaf-cohorts-Z}. 

\begin{table}[H]
    \centering
    \caption{Leaf-cohorts $\Gamma_1,\ldots,\Gamma_L$ from the causal tree $\mathcal{C}_{Z}$} \label{tab: leaf-cohorts-Z}
    \begin{tabular}{cl|cl}
    \hline
    Leaf-cohorts & \multicolumn{1}{c|}{Cohort regions} & Leaf-cohorts & \multicolumn{1}{c}{Cohort regions} \\ \hline
    $\Gamma_1$  & $\{Z \leq 2\}$  &  $\Gamma_4$  & $\{5<Z\leq 7\}$  \\
    $\Gamma_2$  & $\{2<Z\leq 3\}$ &  $\Gamma_5$  & $\{7<Z\leq 10\}$  \\
    $\Gamma_3$  & $\{3<Z\leq 5\}$ &  $\Gamma_6$  & $\{Z>10\}$  \\ 
    \hline
    \end{tabular}
\end{table}

The next step is that, for each heterogeneous cohort $\Pi_i$, $i=1 \cdots K$, we would like to find the estimated causal effects for given treatment values, and report the optimal treatment associated with this heterogeneous cohort. For any $\mathbf{x} \in range(\mathbf{X})$ and $t\in range(T)/\{0\}$, we can identify the unique cohort pair $(\Pi_i, \Gamma_j)$ such that $\mathbf{x}\in\Pi_i$ and $t\in \Gamma_j$. We estimate the treatment effect as $\hat\delta(\mathbf{x}, t) = \hat\delta(\Pi_i, \Gamma_j)$ for any $\mathbf{x}\in\Pi_i$ and $t\in \Gamma_j$. What's more, for each pair of $(\Pi_i, \Gamma_j)$, we can find a unique leaf node $\ell_k$ in $\mathcal{C}_{\mathbf{X}, Z}$, corresponding to a leaf-cohort $\Lambda_k$ such that $\Pi_i \subseteq \Lambda_{k}(\mathbf{X})$ and $\Gamma_j \subseteq \Lambda_{k}(Z)$. The treatment effect can then be estimated as $\hat\delta(\Pi_i, \Gamma_j) = \hat\delta(\Lambda_k)$. Considering the leaf-cohorts in Tables \ref{tab: leaf-cohorts} - \ref{tab: leaf-cohorts-Z}, Table \ref{tab: leaf-cohorts-estimated-causal-effects} provides a summary of relationships between $(\Pi_i, \Gamma_j)$ and $\Lambda_k$.

\begin{table}[ht]
    \centering
    \caption{Estimated causal effects:  $\hat\delta(\Pi_i, \Gamma_j) = \hat\delta(\Lambda_k)$ } \label{tab: leaf-cohorts-estimated-causal-effects}
    \begin{tabular}{c|cccccc}
    \hline
    & \multicolumn{6}{c}{Leaf-cohorts of $\mathcal{C}_{Z}$} \\ \hline
    Leaf-cohorts of $\mathcal{C}_{\mathbf{X}}$ & $\Gamma_1$ & $\Gamma_2$ & $\Gamma_3$ & $\Gamma_4$ & $\Gamma_5$ & $\Gamma_6$ \\ \hline
    $\Pi_1$ & $\Lambda_{1}$  & $\Lambda_{1}$  & $\Lambda_{1}$  & $\Lambda_{4}$  & $\Lambda_{4}$  & $\Lambda_{6}$ \\  
    $\Pi_2$ & $\Lambda_{1}$  & $\Lambda_{1}$  & $\Lambda_{1}$  & $\Lambda_{5}$  & $\Lambda_{5}$  & $\Lambda_{6}$ \\ 
    $\Pi_3$ & $\Lambda_{2}$  & $\Lambda_{2}$  & $\Lambda_{2}$  & $\Lambda_{4}$  & $\Lambda_{4}$  & $\Lambda_{6}$ \\ 
    $\Pi_4$ & $\Lambda_{2}$  & $\Lambda_{2}$  & $\Lambda_{2}$  & $\Lambda_{5}$  & $\Lambda_{5}$  & $\Lambda_{6}$ \\ 
    $\Pi_5$ & $\Lambda_{3}$  & $\Lambda_{3}$  & $\Lambda_{3}$  & $\Lambda_{5}$  & $\Lambda_{5}$  & $\Lambda_{6}$ \\ 
    $\Pi_6$ & $\Lambda_{7}$  & $\Lambda_{7}$  & $\Lambda_{10}$ & $\Lambda_{10}$ & $\Lambda_{10}$ & $\Lambda_{10}$ \\ 
    $\Pi_7$ & $\Lambda_{8}$  & $\Lambda_{9}$  & $\Lambda_{10}$ & $\Lambda_{10}$ & $\Lambda_{10}$ & $\Lambda_{10}$ \\
    $\Pi_8$ & $\Lambda_{11}$ & $\Lambda_{11}$ & $\Lambda_{11}$ & $\Lambda_{11}$ & $\Lambda_{12}$ & $\Lambda_{12}$ \\ 
    $\Pi_9$ & $\Lambda_{13}$ & $\Lambda_{13}$ & $\Lambda_{13}$ & $\Lambda_{13}$ & $\Lambda_{13}$ & $\Lambda_{13}$ \\
    \hline
    \end{tabular}
\end{table}

One attractive feature of our proposed approach is that the task of estimating causal effects can be achieved simultaneously with the tree construction procedures from $\mathcal{C}_{\mathbf{X},Z}$ to $\mathcal{C}_{\mathbf{X}}$.  The information of estimated causal effects has been encoded in the node name of leaf nodes in $\mathcal{C}_{\mathbf{X}}$, in which we keep track of the source leaf nodes in $\mathcal{C}_{\mathbf{X}, Z}$ for the merged nodes in $\mathcal{C}_{\mathbf{X}}$. For example, for the leaf-cohort $\Pi_1$, it is associated with the leaf node $a_{\mathbf{X},1}$ with  {\small $id = 1\cup 4 \cup 6$}. This means in order to estimate the causal effects in $\Pi_1$ with respect to various treatment values, we only need to look into its three source codes, which corresponds to $\Lambda_1$ (Leaf $\ell_1$), $\Lambda_4$ (Leaf $\ell_4$) and $\Lambda_6$ (Leaf $\ell_6$). The optimal treatment for cohort $\Pi_1$ is determined by 
{
\small
\begin{align*}
    t^* = \left\{
    \begin{aligned} 
    & 0 \quad \text{if } \max ( \hat\delta(\Lambda_{1}), \hat\delta(\Lambda_{4}), \hat\delta(\Lambda_{6}) \leq 0, \\ 
    & \underset{t \in range(Z)}{\arg\max}~ \hat{\delta}(\Pi_1, t) = \{ t \in \Lambda_k \text{ for } k \text{ satisfying } \hat\delta(\Lambda_k) = \max ( \hat\delta(\Lambda_{1}), \hat\delta(\Lambda_{4}), \hat\delta(\Lambda_{6}) \} \quad \text{otherwise}.
    \end{aligned} 
    \right.
\end{align*}
}

In summary, this observation enables us to estimate the treatment effects for one cohort in $\mathcal{C}_{\mathbf{X}}$ by only looking at its source nodes in $\mathcal{C}_{\mathbf{X}, Z}$ . This feature greatly facilitates the task of identifying the optimal treatment.

\section{Proof of Theorem \ref{theorem: identifiability}} 
\label{sec:proofs}
Under the assumptions given in Section \ref{section: intro}, we have
{\small $\tau(z, \mathbf{x}) = E[Y \mid \mathbf{X} = \mathbf{x}, T = z] - E[Y \mid \mathbf{X} = \mathbf{x}, T = 0].$}
It follows from the definition of $W$, $Z_1$ and $Z$ that 
{\small $E[Y \mid \mathbf{X} = \mathbf{x}, T = z] = E[Y \mid \mathbf{X} = \mathbf{x}, W = 1, Z_1 = z] = E[Y \mid \mathbf{X} = \mathbf{x}, W = 1, Z = z]. $}
Next, it follows from the independence of $(Z, \mathbf{X})$ and $W$ that
{\small $E[Y \mid \mathbf{X} = \mathbf{x}, W = 1, Z = z] = E[Y(W = 1) \mid \mathbf{X} = \mathbf{x}, Z = z].$} 
Hence, 
{\small 
    $E[Y \mid \mathbf{X} = \mathbf{x}, T = z] = E[Y(W = 1) \mid \mathbf{X} = \mathbf{x}, Z = z].$ 
}

Since $T=0$ is equivalent to $W=0$ (by definition), we have
{\small $E[Y \mid \mathbf{X} = \mathbf{x}, T = 0] = E[Y \mid \mathbf{X} = \mathbf{x}, W = 0] = E[Y(W = 0) \mid \mathbf{X} = \mathbf{x}]. $} 
Finally, it follows from the independence of $Z$ and $(Y(W=0), \mathbf{X})$ that
{\small $E[Y(W = 0) \mid \mathbf{X} = \mathbf{x}] = E[Y(W = 0) \mid \mathbf{X} = \mathbf{x}, Z = z].$}
With the two, we have
{\small
    $E[Y \mid \mathbf{X} = \mathbf{x}, T = 0] = E[Y(W = 0) \mid \mathbf{X} = \mathbf{x}, Z = z].$
}
This completes the proof. 

\section{Discussions on Simulation Designs} \label{sec: more on experiments}
In this section, we introduce the intuition behind our simulation designs. We let $$\eta(x_1, x_2) = -2 + \frac{2}{1+\exp(-12(x_1 - 0.2))} \times \frac{2}{1+\exp(-12(x_2 - 0.2))} $$ denote a function of heterogeneous features $\mathbf{X}_i = (X_{1i}, X_{2i})$. Figure (\ref{fig:heatmap-2:3}) provides a graphical illustration of the values of $\eta(x_1,x_2)$ for $(x_1,x_2)$ within a unit square. Blue and red depict the regions in which $\eta(x_1, x_2)$ takes positive and negative values, respectively. The heterogeneous nature of how the function $\eta(x_1, x_2)$ responds to features makes it the building block for designing our experiments. Besides $\eta(x_1,x_2)$, we include $\eta(x_1,1-x_2)$ to infuse more heterogeneity to the data.
By defining the function of heterogeneous treatment effects, denoted by $f(x_1,x_2, t)$, as a function consisting of $\eta(x_1, x_2)$ and $\eta(x_1, 1-x_2)$, different pairs of $(x_1,x_2)$ will have various optimal treatments. We choose $f(x_1,x_2, t)$ to be the following functions under the different settings: 


\begin{description}[noitemsep]
    \item[(i)] Continuous: 
    {
    \begin{align*}
     f(x_1,x_2,t) &=  5\eta(x_1, x_2) \mathds{1}_{\{t\in(0,0.3]\}} ~-  5\eta(x_1, x_2)\mathds{1}_{\{t\in(0.3,0.5]\}} +  \eta(x_1, 1-x_2)\mathds{1}_{\{t\in(0.5,0.7]\}} \\ &\qquad - \eta(x_1, 1-x_2)\mathds{1}_{\{t\in(0.7,1]\}}.
    \end{align*}
    }
    \item[(ii)] Ordinal:
    {
    $$ f(x_1,x_2,t) =  5\eta(x_1, x_2) \mathds{1}_{\{t=1,2\}} - 5\eta(x_1, x_2)\mathds{1}_{\{t=5\}} +  \eta(x_1, 1-x_2)\mathds{1}_{\{t=3,4\}} - \eta(x_1, 1-x_2)_{\{t=6\}}.$$
    }
    \item[(iii)] Categorical:
    \begin{align*}
    f(x_1,x_2,t) &=  5\eta(x_1, x_2) \mathds{1}_{\{t=a\}} - 5\eta(x_1, x_2) \mathds{1}_{\{t=b\}} +  \eta(x_1, 1-x_2)\mathds{1}_{\{t=c\}} -\eta(x_1, 1-x_2)\mathds{1}_{\{t=d\}}.
    \end{align*}
\end{description}

Using the categorical case as an example, Table \ref{tab:examplefunction} summarizes the values of heterogeneous treatment effects $f(x_1,x_2,t)$ with four different pairs of heterogeneous features $(x_1,x_2)$. From the table we can see that for $(x_1, x_2)$ at different locations in the unit square, it responds distinctively with respect to different treatments. As a result, the optimal treatment varies across various pairs of heterogeneous features.
 
    

    \begin{align*}
        f(x_1,x_2,t) = \left\{ 
        \begin{aligned}
        & 5\eta(x_1, x_2),    & t = a. \\
        & -5 \eta(x_1, x_2),  & t = b. \\
        & \eta(x_1, 1-x_2),   & t = c. \\
        & -\eta(x_1, 1-x_2),  & t = d. \\ 
        & 0,                  & t = 0.
        \end{aligned}
        \right.
    \end{align*}

\begin{table}[ht]
    \centering
    \caption{Values of $f(x_1,x_2,t)$ with different pairs of $(x_1,x_2)$}\label{tab:examplefunction}
    \begin{tabular}{c|cccc|ccc}
     \toprule
    \multirow{2}{*}{$(x_1, x_2)$} &  \multicolumn{4}{c|}{Treatment $t$} & optimal treatment \\
    \cline{2-5}
     & $a$ & $b$ & $c$  & $d$ & for $(x_1,x_2)$ \\ 
     \hline
     $(0.8,0.8)$ &  \textbf{9.970} & -9.970 & -0.001 &  0.001 & $a$ \\
     $(0.2,0.8)$ & -0.007 &  0.007 & -1.000 &  \textbf{1.000} & $d$ \\
     $(0.8,0.2)$ & -0.007 & 0.007 &  \textbf{1.994} & -1.994 & $c$ \\
     $(0.2,0.2)$ & -5.000 &  \textbf{5.000} & -0.001 &  0.001 & $b$\\
    \bottomrule
    \end{tabular}
\end{table}

\end{document}